\documentclass[journal]{IEEEtran}
\usepackage{xcolor,soul,framed} %,caption
\usepackage{subfigure}
\usepackage[numbers,sort&compress]{natbib}
\colorlet{shadecolor}{yellow}
\usepackage[pdftex]{graphicx}
\usepackage{algorithmic}
\usepackage{booktabs}
\graphicspath{{../pdf/}{../jpeg/}}
\DeclareGraphicsExtensions{.pdf,.jpeg,.png,.eps}
\usepackage[cmex10]{amsmath}
\usepackage{array}
\usepackage[pdftex]{graphics}
\usepackage{multirow} %multirow for format of table 
\usepackage{mdwmath}
\usepackage{mathabx}
\usepackage{mdwtab}
\usepackage{eqparbox}
\usepackage{url}
\usepackage{color}
\usepackage{algorithm}
\usepackage[justification=centering]{caption}
\usepackage{caption}
\usepackage{graphicx}
\usepackage{float} 
\usepackage{caption}
\usepackage{graphicx, subfig}
\usepackage{subfigure}
\usepackage{harpoon}
\begin{document}

% \captionsetup[figure]{name={Fig.},labelsep=period} % 将Fig.1：后面的冒号变成点

%  \author{Zhe Wang, ~\IEEEmembership{Student Member,~IEEE,}
 % Yansha Deng,~\IEEEmembership{Senior Member,~IEEE,}\\/
 % and A. Hamid Aghvami, ~\IEEEmembership{Life Fellow,~IEEE}
\bstctlcite{IEEEexample:BSTcontrol}
    \title{Goal-oriented Semantic Communications for Avatar-centric Augmented Reality}
  \author{Zhe Wang, ~\IEEEmembership{Graduate Student Member,~IEEE, }
 Yansha Deng, ~\IEEEmembership{Senior Member,~IEEE, }
 and A. Hamid Aghvami, ~\IEEEmembership{Life Fellow,~IEEE }
 % <-this % stops a space

  % \thanks{Manuscript received July 10, 2023. \hl{This paper is an expanded paper from the IEEE Globecom Symposium held on December 17-22, 2023 in Kuala Lumpur, Malaysia} 
  % % This work was funded in part by the Office of Naval Research under the Defense Advanced Research Projects Agency (DARPA) Microscale Power Conversion (MPC) Program under Grant N00014-11-1-0931, and in part by the Advanced Research Projects Agency-Energy (ARPA-E), U.S. Department of Energy, under Award Number DE-AR0000216.
  % }
  \thanks{This work was supported in part by UKRI under the UK government’s Horizon Europe funding guarantee (grant number 10061781), as part of the European Commission-funded collaborative project VERGE, under SNS JU program (grant number 101096034). This work is also a contribution by Project REASON, a UK Government funded project under the FONRC sponsored by the DSIT. An earlier version of this paper was presented in part at the
2023 IEEE Global Communications Conference in December 2023 [DOI:
10.1109/GLOBECOM54140.2023.10437075].
  
  Z. Wang, Y. Deng, and A. Hamid Aghvami (Emeritus Professor) are with the Department of Engineering, King’s College
London, Strand, London WC2R 2LS, U.K. (e-mail: tylor.wang@kcl.ac.uk;
yansha.deng@kcl.ac.uk; hamid.aghvami@kcl.ac.uk) (Corresponding author: Yansha Deng).
Youtube: https://youtu.be/n9YPF979m\_0. Github:https://github.com/kcl-yansha/PCDataset

}% <-this % stops a space
%   \thanks{T. Reveyrand is with the XLIM Laboratory, UMR 7252, University of Limoges, 87060 Limoges, France (e-mail: tibault.reveyrand@xlim.fr).}%
%   \thanks{I. Ramos and Z. Popovic are with the Department of Electrical, Computer and Energy Engineering, University of Colorado, Boulder, CO, 80309-0425 USA (e-mail: ignacio.ramos@colorado.edu; zoya.popovic@colorado.edu).}% <-this % stops a space
%   \thanks{E. Falkenstein is with Qualcomm Inc., 6150 Lookout Road
% Boulder, CO 80301 USA (e-mail: erez.falkenstein@gmail.com).}
}

% \markboth{IEEE Transactions on Wireless Communications, VOL.~60, NO.~12, DECEMBER~2023
% }
% {Roberg \MakeLowercase{\textit{et al.}}: Task-oriented and semantics-aware Communication Framework for Augmented Reality}

% ====================================================================
\maketitle
% \newcommand\blfootnote[1]{%
% \begingroup
% \renewcommand\thefootnote{}\footnote{#1}%
% \addtocounter{footnote}{-1}%
% \endgroup
% }

% === ABSTRACT ====================================================================
% =================================================================================
\begin{abstract}
%\boldmath
% Upon the arrival of the emerging metaverse and its related applications in Augmented Reality (AR), the current bit-oriented network struggles to support real-time changes for the vast amount of associated information, which hinders its development. A critical revolution in Sixth Generation (6G) networks is envisioned through the joint exploitation of data context and its importance to the task, leading to a communication paradigm shift toward semantics and effectiveness levels. However, no explicit and systematic communication framework has been proposed for AR to incorporate these levels. To address this problem, we develop a novel task-oriented, semantics-aware communication framework for augmented reality (TSAR) to enhance communication efficiency and effectiveness in potential 6G scenarios. Specifically, we first analyze the unique components of moving avatars and summarize the semantics information of avatars along with the end-to-end transmission of the framework. Next, we propose a detailed TSAR communication framework, including TSAR  information, semantics encoder, task-oriented execution, semantics decoder, and effectiveness level performance metrics. Finally, numerous experiments demonstrate that compared to the conventional cloud point transmission framework, our proposed task-oriented and semantics-aware communication framework significantly reduces transmission delay while improving video recovery accuracy in geometry and color aspects by up to 82.4\% and 20.4\%, respectively.
\textcolor{black}{With the emergence of the metaverse and its applications in representing humans and intelligent entities in social and related augmented reality (AR) applications. The current bit-oriented network faces challenges in supporting real-time updates for the vast amount of associated information, which hinders development.}
% Upon the advent of the emerging metaverse and its related applications in Augmented Reality (AR), the current bit-oriented network struggles to support real-time changes for the vast amount of associated information, hindering its development. 
Thus, a critical revolution in the sixth generation (6G) networks is envisioned through the joint exploitation of information context and its importance to the goal, leading to a communication paradigm shift towards semantic and effectiveness levels. However, current research has not yet proposed any explicit and systematic communication framework for AR applications that incorporate these two levels. To fill this research gap, this paper presents a goal-oriented semantic communication framework for augmented reality (GSAR) to enhance communication efficiency and effectiveness in 6G. Specifically, we first analyse the traditional wireless AR point cloud communication framework and then summarize our proposed semantic information along with the end-to-end wireless communication. We then detail the design blocks of the GSAR framework, 
% , which includes semantic information extraction with deep learning, avatar pose recovery, base knowledge selection, and task-oriented semantics-aware wireless communication, 
covering both semantic and effectiveness levels. Finally, numerous experiments have been conducted to demonstrate that, compared to the traditional point cloud communication framework, our proposed GSAR significantly reduces wireless AR application transmission latency by 95.6\%, while improving communication effectiveness in geometry and color aspects by up to 82.4\% and 20.4\%, respectively.
\end{abstract}

% === KEYWORDS ====================================================================
% =================================================================================
% \begin{IEEEkeywords}
% {Metaverse, augmented reality, semantics communication, semantics coding, cloud point, end-to-end communication}
% \end{IEEEkeywords}

% \fi
\IEEEpeerreviewmaketitle

% ====================================================================
% ====================================================================
% ====================================================================

% \cite{pacchioni2020virtual}

% \cite{garzon2019systematic}

% === I. INTRODUCTION =============================================================
% =================================================================================
\section{Introduction}
\IEEEPARstart{T}{he} metaverse, as an expansion of the digital universe, has the potential to significantly influence people's lives, affecting their entertainment experiences and social behaviors. 
Specific applications such as augmented reality (AR), virtual reality (VR), and other immersive technologies within the metaverse have demonstrated remarkable potential in various areas, including virtual conferences, online education, and real-time interactive games, capturing the attention of both industry and academia \cite{ning2021survey}.
% \cite{pacchioni2020virtual, garzon2019systematic, ning2021survey}.
These applications, also referred as extended reality (XR), need to process rich and complex data, such as animated avatars, point cloud, and model mesh, to create immersive experiences for clients \cite{park2022metaverse}. However, the extensive transmission of information and high bandwidth requirements within the XR pose significant challenges for its wider applications, particularly in avatar-related applications that necessitate real-time client communication and interaction. 
\textcolor{black}{The existing communication networks fails to achieve such high bandwidth requirement and thus can not adequately support XR applications, necessitating the development of 6G technology to enhance its applications for further advancement \cite{hu2020cellular, hu2021vision}.}
Specifically, to ensure a good quality of experience (QoE) in AR applications, a transmission latency of less than 20 ms is required, which is 20 times less than the transmission latency tolerated in video communication applications \cite{van2020human}. 
Due to the nature of numerous sensing data in AR applications, more packets need to be transmitted in such a short time, which consequently increases the demand for bandwidth.
\textcolor{black}{Previous research has demonstrated that metaverse clients typically require a bandwidth of about 5.6 Gbps for raw metaverse data communication \cite{dong2022metaverse}. However, the average 5G global wireless download bandwidth is only about 160 Mbps, posing significant challenges for the development of metaverse applications \cite{kumar20235g}.}
\textcolor{black}{This results in challenges in meeting the transmission latency and bandwidth requirements, particularly for the transmission of more complex and comprehensive data in AR services \cite{liao2022kitti}, underscores the urgency for further research in communication technology. Such advancements are crucial to facilitate real-time, immersive experiences in AR-based applications.}
% This growing concern about the transmission latency and bandwidth in AR application services highlights the need for further research in communication technology to ensure a real-time immersive experiences for clients in AR-related applications.

To address the high bandwidth requirements for diplomas in wireless communication in AR applications, the concept of semantic communications has been proposed \cite{kountouris2021semantics}. This approach aims to facilitate communication at the semantic level by exploring not only the content of traditional text and speech data but also the information's freshness. Initial research on semantic communications for text \cite{yan2022resource}, speech \cite{weng2021semantic}, and image data \cite{jiang2022wireless} has mainly focused on identifying the semantic content of traditional data. Other research in semantic communications on sensor and control data emphasizes the goal requirements for using information freshness, such as age of information (AoI) \cite{maatouk2022age}, as a semantic metric to estimate timeliness and evaluate the importance of the information.
It should be noted that these AoI-related semantic communications approaches are unable to adequately capture the importance of specific information with inherent importance in the emerging AR dataset. 
% This highlights the need to develop new strategies and techniques that effectively incorporate semantic communication into AR, considering not only the timeliness of information but also its relevance and sufficiency for a given application. In \cite{zhou2022task}, a generic task-oriented and semantics-aware communication framework is envisioned, taking into account designs at both the semantic and effectiveness levels for various tasks with diverse data types. Another research highlights the importance of task-oriented and semantics-aware communication in robotic control (TSRC). This research notably exploits the context of data, emphasizing its critical role in successful task execution at both the transmitter and receiver ends. Although task-oriented performance has been implemented in various aspects, a specific and concrete task-oriented, semantics-aware communication framework for AR applications to improve task performance has not yet been proposed.
\textcolor{black}{This highlights the need to develop new strategies and techniques that effectively incorporate specific goal with semantic communications into AR, considering not only the timeliness of information but also its relevance and sufficiency for a given application. In \cite{zhou2022task}, a generic goal-oriented semantic communication framework is envisioned for robotic applications, taking into account designs at both the semantic and effectiveness levels for various goals with diverse data types. Then, in \cite{wu2023task}, researchers highlight the importance of goal-oriented semantic communications in robotic control (GSRC) by exploiting the context of data, emphasizing its critical role in successful goal execution at both the transmitter and receiver ends. However, although goal-oriented performance has been implemented in these robotic aspects, a specific and concrete goal-oriented semantic communication framework for avatar-centric AR applications to improve goal performance has not yet been proposed.
}

Current XR-related application research typically requires users to utilize head-mounted displays (HMD) \cite{du2022optimal}. These applications generally focus on avatar-centric services, where the use of avatar animation in replacement of real human figures can decrease HMD computing requirements, reduce transmission data, and protect user privacy \cite{fernandez2022life}. This avatar representation method has been implemented in social media platforms, such as TikTok and Instagram, where avatar characters is used for augmented reality video effects. Interestingly, using avatars instead of human has shown no significant differences in social behavior transmission and can even attract users to complete their goal more quickly in gaming situations \cite{pauw2022avatar}. For instance, fitness coaches can employ virtual avatars for AR conferencing to guide training. Games, like Pokémon Go, use avatars in mixed reality to encourage gamer interaction \cite{lemmens2023caught}. 
\textcolor{black}{Avatar-based communication has been considered in \cite{da2019point}, where the point cloud of avatars, structures, and models are transmitted between transmitter and receiver. Goal-related effectiveness level performance metrics, including point-to-point \cite{yang2022no}, peak signal-to-noise ratio for the luminance component \cite{lazzarotto2022influence}, mean per joint position error \cite{pavllo20193d} have been considered to assess the goal achievement of telepresence \cite{yu2021avatars}, point cloud video displaying \cite{xu2021epes}, and avatar pose recovery \cite{liu2022deep}, respectively.} 
\textcolor{black}{
Based on these goals, recent research has also proposed implementing avatar representations such as point clouds, skeleton, and $360^\circ$ images \cite{aseeri2021influence, yu2021avatars}. Although these studies emphasize data extraction from a graphical perspective, wireless AR-related communication applications have not fully addressed the issue of avatar transmission effectiveness from a wireless communication perspective. Furthermore, the bandwidth requirements for such applications remain high.
}
\textcolor{black}{
Users continue to experience suboptimal and lagging AR experiences in areas with moderate signal strength. This suggests that the current AR communication framework has limitations, particularly in identifying a better method for avatar representation to enhance communication. 
Specifically, there is a demand for approaches that require less bandwidth and improve goal performance, which must be addressed \cite{haruna2023augmented}.}
% Users continue to experience suboptimal and lagging AR experiences in areas with moderate signal strength, indicating that the current AR communication framework has limitations, particularly in identifying a better avatar representation method for more effective communication, which need to be addressed.

Several studies have recently begun to explore the representation of avatars in wired communication. Different data types have been designed to represent avatars, which results in diverse avatar reconstruction required at the client side and limited transmission effectiveness evaluation capabilities for AR. 
For instance, skeleton elements have been proposed as a means to represent avatars, where motion capture devices are used to record skeletal positions. The recorded avatar movements are then replayed in wired HMDs, and the differences in skeleton position between transmitter and receiver are measured to evaluate wired AR communication \cite{wu2019towards}. 
% Meanwhile, in \cite{van2020objective}, 2D videos captured from various viewing angles are used to represent 3D avatars in wired transmission scenarios. These videos are subsequently employed in the reconstruction of 3D avatars after transmission, with the avatar pose differences between the wired transmitter and receiver used as the metric for evaluation.}
% However, how to best extract semantic information and evaluate the avatar-centric display task is still unclear in a wireless communication AR application.
However, how to best extract semantic information that reflects the importance and context of information related to the avatar-centric displaying is still unclear in a wireless communication AR application.
The presence of redundant messaging can lead to an increase in transmission packets, resulting in decreased efficiency of wireless communication and ultimately impacting the user's viewing experience.
Inspired by the 3D keypoints extraction method presented in \cite{you2020keypointnet}, we propose a goal-oriented semantic communication framework in AR (GSAR)  for avatar-centric end-to-end AR communication. In contrast to traditional point cloud AR communication frameworks that rely solely on point cloud input, our proposed GSAR extracts and transmits only essential semantic information.
% , significantly reducing the communication bandwidth required. This research is the first to investigate and evaluate the AR wireless communication framework considering task-oriented and semantics-aware levels using real 3D scenery datasets. 
To the best of our knowledge, our contributions can be summarized as follows:
\begin{enumerate}
\item \textcolor{black}{We propose a goal-oriented semantic communication framework in augmented reality (GSAR) for \textcolor{black}{interactive avatar-centric displaying} applications with an integration of the semantic and effectiveness levels design, which includes semantic information extraction, goal-oriented semantic wireless communications, avatar pose recovery and rendering.}

\item \textcolor{black}{
We apply an avatar-based semantic ranking (AbSR) algorithm to extract features from the avatar skeleton graph using shared base knowledge and to sort the importance of different semantic information. Additionally, by utilizing channel state information (CSI) feedback, we demonstrate the effectiveness of AbSR in improving avatar transmission quality in wireless AR communication.}
% \item \textcolor{black}{We apply an avatar-based semantic ranking (AbSR) algorithm to abstract features from the avatar skeleton graph using shared base knowledge and measure the importance of different semantic information. By utilizing Channel State Information (CSI) feedback, the AbSR can improve the avatar transmission quality in the wireless AR communication.}

\item \textcolor{black}{We have conducted a series of experiments comparing our proposed GSAR framework with the traditional point cloud communication framework. 
% These experiments incoperate semantics information extraction, task-oriented wireless communication, and avatar reconstruction. 
Our results indicate that our proposed GSAR framework outperforms the traditional point cloud communication framework in terms of color quality, geometry quality, and transmission latency for avatar-centric displayinging, with improvements of up to 20.4\%, 82.4\% and 95.6\% respectively.}
\end{enumerate}

The rest of the paper is organized as follows: In section II, we present the system model and problem formation, covering both the traditional point cloud and the GSAR frameworks. Section III details the design principles for semantic level. Section IV details the design principles for effectiveness level. Section V demonstrates the avatar movement and experimental performance evaluation. Finally, Section VI concludes this paper.
% === II. Harmonically-Terminated Power Rectifier Analysis ========================
% =================================================================================
% \begin{figure*}[t]
% \centering
% \includegraphics[width=\textwidth]{Traditional_Communication.png}
% % [height=4.5cm]表示高度
% %[width=9.5cm]表示宽度
% %{111.eps}表示eps格式的图片，名为111
% \caption{semantics encoder training under different backbone networks}
% %图片的名称
% \label{Fig_2Training_results}
% %图片的标签，用于文章中的引用，注意到标签的数字与实际文章显示的数字可能不同
% \end{figure*}

% \begin{figure*}[t]
%   \begin{center}
%   \includegraphics[width=\textwidth]{Traditional_Communication.png}\\
%   \caption{Microwave rectifier circuit diagram. An ideal blocking capacitor $C_b$ provides DC isolation between the microwave source and rectifying element.  An ideal choke inductor $L_c$ isolates the DC load $R_{DC}$ from RF power.}
%   \end{center}
%   \label{TranditionalFramework}
% \end{figure*}

\section{System Model and Problem Formation}
% In this section, we begin by outlining the existing traditional point cloud communication framework used in AR applications. We then provide a detailed description of our proposed TSAR framework, which considers not only the bit-level, but also the semantics and task levels. Finally, we present the problem formulation and the objective function.
In this section, we first describe the existing traditional point cloud communication framework for AR applications. Then, we present our wireless communication channel model implemented in both the point cloud communication framework and the GSAR.
We further introduce our proposed GSAR in detail, which considers not only the bit-level but also the semantic and effectiveness levels.  Finally, we present the problem formation and the objective function.
% =======
% FIG. 01
% =======
% \begin{figure}
%   \begin{center}
%   \includegraphics[width=3.5in]{pdf/01.pdf}\\
%   \caption{Microwave rectifier circuit diagram. An ideal blocking capacitor $C_b$ provides DC isolation between the microwave source and rectifying element.  An ideal choke inductor $L_c$ isolates the DC load $R_{DC}$ from RF power.}\label{circuit_diagram}
%   \end{center}
% \end{figure}
%  \begin{figure}[t]
%   \centering
%   \includegraphics[width=8.8cm]{TraditionalHarlf.png} %1.png是图片文件的相对路径
%   %\includegraphics[width=.8\textwidth]{Figure\1.png}
%   \caption{Traditional point cloud communication framework} %caption是图片的标题
%   \label{traditionalframework} %此处的label相当于一个图片的专属标志，目的是方便上下文的引用
% \end{figure}
 \begin{figure}[t]
  \centering
  \includegraphics[width=8.8cm]{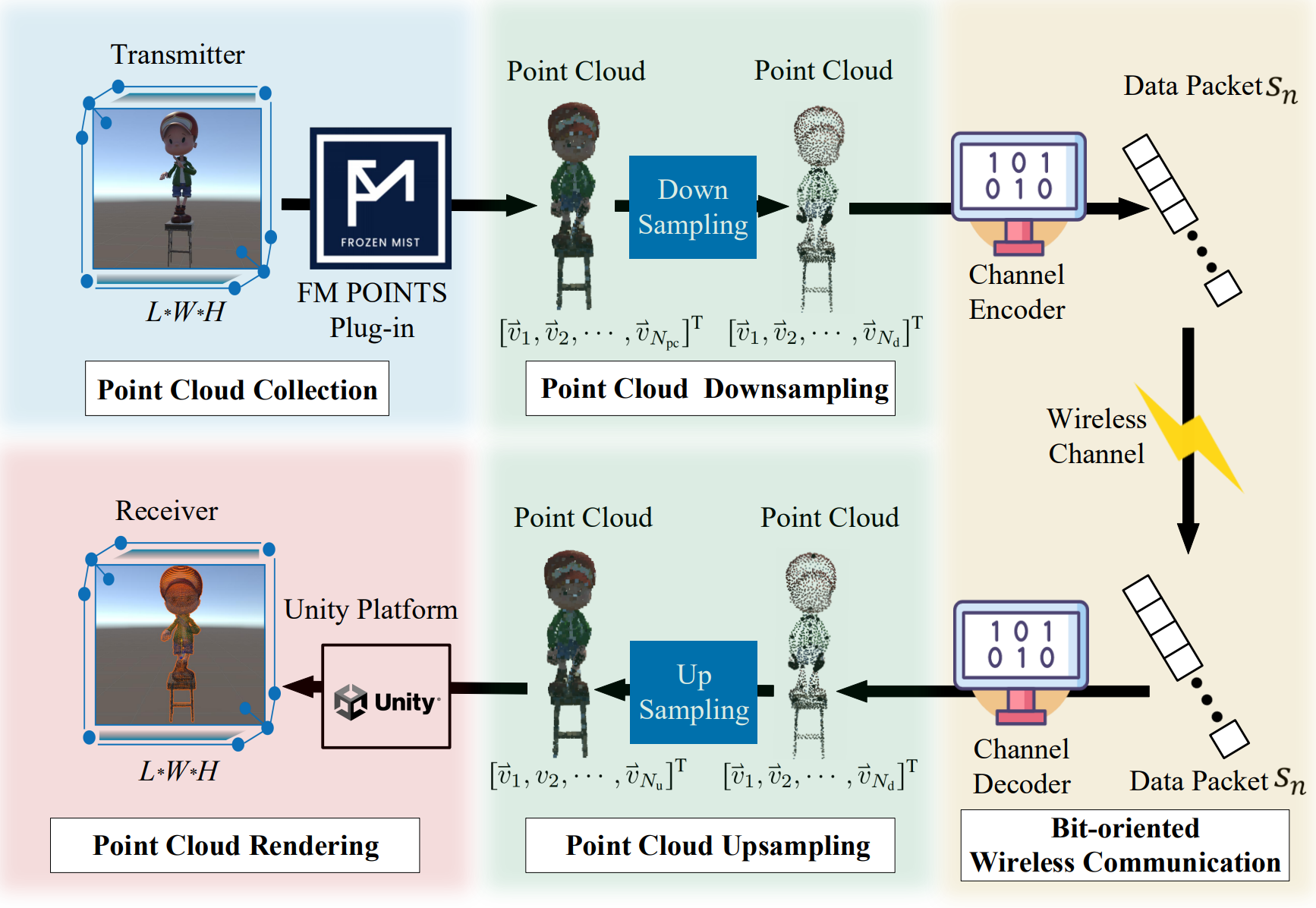} %1.png是图片文件的相对路径
  \caption{Traditional point cloud communication framework} %caption是图片的标题
  \label{traditionalframework} %此处的label相当于一个图片的专属标志，目的是方便上下文的引用
\end{figure}
\subsection {Traditional Point Cloud Communication Framework}
As shown in Fig. \ref{traditionalframework}, the procedures for traditional point cloud communication in AR applications typically consist of point cloud collection, downsampling, upsampling, and rendering.

\subsubsection {Point Cloud Collection}
% We consider an avatar-centric conferencing and gaming application within the AR domain, which involves transmitting avatar animations to the client user for viewing on an HMD in a 3D area with length $L$, height $H$, and width $W$. This is currently a prominent research area in the metaverse field\cite{fernandez2022life}. To ensure accurate transmission of avatar $A_i$ in an AR application, high-resolution point clouds $P^{A_i}$ of the avatar must be captured and transmitted to the client side.
% Generating large amounts of sensor data from the Unity3D platform using plugins has been a popular research focus among developers, with examples such as FMPC and Headbox\cite{volonte2022headbox}. FM POINTS is a comprehensive point cloud visualization plugin for Unity3D that can convert any 3D object in Unity3D scenes into real-time point clouds. The generated point clouds can be represented as $P^{A_i}=\{V^{A_i}_1, ..., V^{A_i}_{N_{pc}}\}$, where $N_{pc}$ denotes the number of point clouds generated. Generally, every frame requires over 1500 thousand point clouds to achieve acceptable Quality of Service (QoS) for clients\cite{zhang2021towards}. The data of each point $V$ contains six values, representing the 3D dimensional location $(l_x, l_y, l_z)$ and an RGB color representation $(c_r, c_g, c_b)$.
We focus on interactive avatar-centric displaying and gaming AR applications, which are promising applications in the metaverse \cite{fernandez2022life}. These AR applications require transmitting avatar animations and other stationary background models to the client side for displaying on an HMD in the area with dimensions length $L$, height $H$, and width $W$.
To guarantee a smooth viewing experience of the AR scenery at the client side, high-resolution point cloud of both the moving avatar and stationary background models need to be captured and transmitted to the client side. Current Unity3D platform have numerous plugins for generating sensor data in real time, such as FM POINTS, which is a comprehensive point cloud visualization plugin that can transform the whole AR scenery or any 3D models into real-time point cloud. The information for each point $\overrightharp{v}_{i}$ can be represented as
% comprises six parameters, including 
\begin{equation}
\overrightharp{v}_{i}=(\overrightharp{l}_i, \overrightharp{c}_i)=(l_\text{x}, l_\text{y}, l_\text{z}, c_\text{r}, c_\text{g}, c_\text{b}),
\end{equation}
where the $\overrightharp{l}_i$ and $\overrightharp{c}_i$ represent the three-dimensional location and RGB color of point, respectively. The generated point cloud $\mathbf{P}_\text{pc}$ of the whole AR scenery consist of thousands of points $v_i$, which can be represented as
% overrightharp
 % ($l_\text{x}, l_\text{y}, l_\text{z}$)  ($c_\text{r}, c_\text{g}, c_\text{b}$)
% $P^{A_i}=\{v^{A_i}_1, ..., v^{A_i}_{N_{pc}}\}$, 
% \begin{equation}
% P_{\text{ar}}={[v_\text{1},v_\text{2} ..., v_{\text{N}_{\text{pc}}}]},
% \end{equation}
\begin{equation}
\mathbf{P}_{\text{pc}}=[\overrightharp{v}_{1} , \overrightharp{v}_{2} , \cdots , \overrightharp{v}_{{N}_{\text{pc}}}]^\text{T},
\end{equation}
where ${N}_{\text{pc}}$ denotes the total number of generated point cloud of AR scenery. 
\textcolor{black}{
Typically, each 3D object needs to be represented by over 1,500 thousand point cloud in each frame to achieve a satisfactory viewing experience for clients \cite{zhang2021towards}, and each frame necessitates the transmission of 1.5 Gb of graphics data for multiple virtual objects scenery construction \cite{mehrabi2021multi}. 
}

\subsubsection {Point Cloud Downsampling and Upsampling}
% Before the sensed data is captured on the server side and after the sensed data is transmitted, a downsample and upsample algorithm will be conducted to maximize the compression of the point cloud and recover the point cloud. Current research on point cloud downsample algorithms includes uniform down sample, random down sample, and voxel down sample. The process of point cloud downsample, denoted as D(.), can be expressed as:
In the traditional point cloud wireless communication framework, the transmission of a large number of point cloud can lead to data congestion at the wireless channel, causing intolerable delays and thus hinders AR application development \cite{huang2022iscom}. To minimize transmission delays, current research explores the use of compression algorithms in point cloud transmission \cite{nardo2022point}. By introducing an downsample algorithm at the transmitter and an upsample algorithm at the receiver, the transmission latency can be reduced through transmitting only the compressed point cloud. \textcolor{black}{The farthest point sampling algorithm \cite{qi2017pointnet++} is utilized as the downsample method, which enables the selection of representative points from the original point cloud while maintaining the overall features of the 3D objects. This algorithm reduces the number of points to be transmitted, thus improving the efficiency of the communication system.}
% Current point cloud downsample algorithms include uniform downsampling, random downsampling, and voxel downsampling \cite{zong2022improved}. 
The process of farthest point downsampling $\mathcal{D(\cdot)}$, can be expressed as
% In a traditional AR wireless communication framework, the processing and transmission of a large number of point clouds consume significant time, which hinders the development of AR applications and results in poor QoS for clients\cite{huang2022iscom}. To minimize transmission delays, current researchers are exploring the use of compression algorithms in point cloud transmission. By applying a downsample algorithm before the transmitter and an upsample algorithm after the receiver, point cloud compression can be maximized and point cloud recovery can be facilitated. Current research on point cloud downsample algorithms includes uniform downsampling, random downsampling, and voxel downsampling\cite{zong2022improved}. The process of point cloud downsampling, denoted as $D(\cdot)$, can be expressed as:
\begin{equation}
\mathbf{P}_{\text{dpc}}=[\overrightharp{v}_{1}, \overrightharp{v}_{2}, \cdots, \overrightharp{v}_{{N}_{\text{d}}} ]^\text{T}=\mathcal{D}(\mathbf{P}_{\text{pc}}),
\end{equation}
% \begin{equation}
% {D_{\text{pc}}}=D(P_{\text{ar}})={[v_\text{1},v_\text{2} ..., {v}_{\text{N}_{\text{d}}}]},
% \end{equation}
% where P a is the point cloud data after compression, and N is the number of points waiting for transmission. The upsample algorithm involves taking a sparse, noisy, and non-uniform point cloud and upsampling it to generate a dense, complete, and uniform point cloud, which remains a challenging research direction. Interpolation algorithms are currently the most robust methods compared to other techniques such as GAN and neural network methods. The process of upsampling, denoted as U(.), can be expressed as:
% Where ${D_{pc}^{A_i}}$ is the point cloud data of avatar $A_i$ after downsampling, $\hat{v}$ refers to a single point cloud, and $N_{d}$ is the number of downsampled points awaiting transmission. Since the physical channel is unstable, the upsampling algorithm aims to convert a sparse, irregular, and non-uniform point cloud into a dense, complete, and uniform one at the receiver side, which remains a challenging research area. Currently, interpolation methods and a variety of deep learning techniques are being explored and implemented for point cloud upsampling. The upsampling process, denoted as $U(\cdot)$, can be expressed as follows:
where ${\mathbf{P}_{\text{dpc}}}$ represents the downsampled point cloud data awaiting transmission, and ${N}_{\text{d}}$ is the total number of downsampled point cloud data. 
Then, the client's view experience can be enhanced by employing an upsampling algorithm for high-resolution point cloud recovery. 
Due to the instability of the wireless channel, the receiver faces the challenge of converting a sparse, irregular, and non-uniform point cloud into a dense, complete, and uniform one. 
% Addressing this issue remains a challenging research area \cite{akhtar2022pu}. 
To address this challenging issue \cite{akhtar2022pu},
\textcolor{black}{the linear interpolation algorithm \cite{chen2020pointmixup} is introduced for the point cloud upsampling process. This algorithm involves estimating the positions of the missing points based on the positions of their neighbors, effectively generating a denser point cloud that closely resembles the original point cloud structure.}
% researchers are investigating and implementing interpolation methods and various Deep Learning (DL) techniques for point cloud upsampling. 
The point cloud upsampling process, denoted as $\mathcal{U}(\cdot)$, can be expressed as
\begin{equation}
{\mathbf{P}_\text{upc}}=[ \overrightharp{v}_{1}, \overrightharp{v}_{2}, \cdots, \overrightharp{v}_{{N}_{\text{u}}} ]^\text{T}=\mathcal{U}({\mathbf{P}_\text{dpc}^{'}}),
\end{equation}
% \begin{equation}
% \mathbf{R_{\text{pc}}}=U(\mathbf{D_{\text{pc}}^{'}})=\begin{bmatrix} v_{1} \ v_{2} \ \vdots \ v_{N_{u}} \end{bmatrix},
% \end{equation}
% Where $R^{'}_{A_i}$ is the reconstructed point cloud after upsampling, and $D^{'}_{A_i}$ is the received point cloud data $D_{A_i}$ following transmission through the physical channel.
where ${\mathbf{P}_{\text{upc}}}$ is the reconstructed point cloud after upsampling, ${N}_{\text{u}}$ represents the total number of upsampled point cloud, and  ${\mathbf{P}_{\text{dpc}}^{'}}$ is the received point cloud data after transmitting ${\mathbf{P}_{\text{dpc}}}$ over wireless channels.
% following transmission through the physical channel, originating from the transmitted data ${\mathbf{P}_{\text{dpc}}}$. 
The upsampling process aims to accurately reconstruct the original point cloud, ensuring that the client-side viewing experience is maintained at a high quality despite the data compression and transmission through an unstable wireless channel.

\subsubsection {Point Cloud Rendering}

\textcolor{black}{
The point cloud rendering process begins when all the ${N}_\text{u}$ point clouds for the AR scenery are received and upsampled. This process prepares the point cloud data for the Unity3D platform and facilitates high-resolution rendering.
The rendering process needs to create a comprehensive $360^{\circ}$ view of the avatar, along with immersive background scenery, which involves point cloud preparation and procedures:
\begin{itemize}
\item[(1)] Point cloud preparation: Point cloud preparation involves formatting points from the received point cloud data. Each point contains information such as three-dimensional location and RGB color value, which determines the point's position and visual depiction within the virtual environment.
% \item[(2)] Point Cloud Processing: Point Cloud Processing includes mesh reconstruction and positioning. It begins with the conversion of these points into a mesh format that is compatible with the Unity3D platform, followed by integration into the AR scene. The positioning process involves translating, rotating, and scaling the point cloud data to ensure accurate alignment within the AR environment.
% \item[(2)] Point cloud processing: Point cloud processing encompasses mesh reconstruction and positioning. It begins with converting these points into a mesh format that is compatible with the Unity3D platform. Then, the Shader, as a specialized program, is used in the rendering process to  manage the levels of lighting, darkness, and color in the virtual environment.  Finally, the positioning phase is implemented to adjust the displaying, which includes translation, rotation, and scaling.
\item [(2)] Point cloud processing: The procedure of point cloud processing includes mesh reconstruction along with positioning. It commences with the transformation of these discrete points into a compatible mesh format for the Unity3D platform. Subsequently, the Shader, a uniquely designed program, is employed during the rendering process to regulate the gradients of illumination, obscurity, and chromaticity within the virtual environment. The final step of this process involves implementing the positioning phase to optimize the visualization, encompassing translation, rotation, and scaling elements. Concurrently, the level of detail (LoD) strategy is invoked in the whole processing process, which dynamically modulates the complexity of a 3D model representation contingent upon its spatial relation to the clients. It renders fewer points when clients are distant and, conversely, more points as they step closer, thereby providing a better viewing experience.
% , to positioning involves translating, rotating, and scaling the point cloud data to ensure accurate alignment within the AR environment.
\end{itemize}
}

\subsection {Wireless Channel Model}
\textcolor{black}{
In the described wireless communication model, which utilizes a frequency-division multiplexing (FDM) scheme within a Rayleigh fading channel affected by additive white Gaussian noise, the division of the wireless channel into multiple parallel subchannels. Each subchannel experiences unique frequency-selective responses due to the Rayleigh fading environment. This results in varying levels of channel gain across different frequencies, leading to different signal-to-noise ratios (SNRs) for each subchannel. The frequency-selective fading, characteristic of such environments, affects the subchannels differently, leading to a range of channel responses and necessitating adaptive strategies for efficient communication.}
% The wireless communication model is characterized by a Rayleigh fading channel, impacted by additive white Gaussian noise and utilizing an Orthogonal Frequency-division Multiplexing (OFDM) scheme. The OFDM approach divides the wireless channel into multiple parallel subchannels. Each subchannel experiences varying levels of noise, leading to different Signal-to-Noise Ratios (SNRs).

% Before wireless transmission, Binary Phase-shift Keying (BPSK), a widely adopted modulation technique, transforms analog signals into digital bits. It accomplishes this by modifying the phase of a carrier signal in response to the data values inputted into the system. Once BPSK processing is complete, \textcolor{black}{the each resulting bit, denoted as $s_n$,} are ready for transmission.
% The multi-path channel within the OFDM scheme, represented as $\mathbf{H}_\text{c}$, can be described as

The wireless communication process begins with source encoding, transforming the awaiting transmit data into the bitstream. Following this, a standard channel encoding is implemented to inject redundancy into the data to be transmitted, safeguarding data integrity and enabling the correction of potential errors during transmission. Traditional communication coding methods, such as turbo coding and low-density parity-check coding, can be utilized in the channel coding process \cite{egilmez2019development}. The encoded bits generated by channel encoding are then carried forward as $b_n$.
Following channel encoding, we implement binary phase-shift keying (BPSK), a widely used modulation technique. BPSK alters the phase of a carrier signal based on the encoded bits $b_n$, resulting in modulated signals denoted as $s_n$. 
\textcolor{black}{
% After channel encoding in digital communication systems, BPSK is a commonly implemented modulation method.
% This modulation technique encodes bits, denoted as $b_n$, resulting in modulated signals represented as $s_n$. 
The modulation process of BPSK can be expressed by the following equation:
\begin{equation}
s_n = 
\begin{cases}
+1, & \text{if } b_n = 1, \\
-1, & \text{if } b_n = 0.
\end{cases}
\end{equation}
where $b_n$ represents the binary input, and $s_n$ represents the modulated output signal. 
In BPSK, a phase shift in the carrier signal is used to convey information. Specifically, the binary input $b_n$ in $\{0,1\}$ is shifted to the output signal $s_n$ in $\{-1,+1\}$. This modulation technique results in two distinct phases of the carrier signal. 
% The simplicity of this representation makes BPSK an efficient and straightforward modulation technique. Despite its relative bandwidth inefficiency compared to more complex schemes, BPSK's resilience against noise and interference secures its position as a foundational and widely-used technique in digital communication.
% In BPSK, a phase shift in the carrier signal is used to convey information. Specifically, the binary input $b_n$ in $\{0,1\}$ is shifted to the output signal $s_n$ in $\{-1,+1\}$. This modulation technique results in two distinct phases of the carrier signal, effectively encoding binary data.
% % BPSK is categorized as a form of binary keying. In binary keying, two discrete signal levels or states symbolize the binary values 0 and 1. 
% This simplicity of representation makes BPSK an efficient and straightforward modulation technique. Despite its relative bandwidth inefficiency compared to more complex schemes, BPSK's resilience against noise and interference secures its position as a foundational and widely-used technique in digital communication.
}

Finally, we take into account the multi-path channel within the FDM, represented as $\overrightharp{H}_\text{c}$. In the wireless channel, each modulated bit $s_n$ is allocated to a subchannel, denoted as $h_n$, and is then ready for transmission over that subchannel. This approach allows for the simultaneous transmission of multiple modulated bits over different subchannels, the channel gains in wireless subchannel is represented as
% thereby enhancing the overall efficiency and robustness of our communication system.
\begin{equation}
\label{snrequ}
\overrightharp{H}_\text{c}=[h_1,h_2, \cdots, h_{N_\text{c}}]^\text{T},
\end{equation}
where $N_\text{c}$ stands for the total number of subchannels in $\overrightharp{H}_\text{c}$, and $h_i$ signifies the channel gain of the $i$-th subchannel. Thus the SNR in each subchannel $h_i$ can be expressed as
\begin{equation}
\mathrm{SNR}_i=\frac{P_i\left\|h_i\right\|^2}{\textcolor{black}{\sigma^2}},
\end{equation}
\textcolor{black}{where $P_i$ represents the transmit signal power}, $\left\|\boldsymbol{h}_i\right\|^2$ denotes the squared norm of the channel gain from the $i$-th subchannel to the destination, \textcolor{black}{and $\sigma^2$ is the noise power.}

Considering the characteristics of each subchannel, the cumulative SNR of the communication process within channel $\overrightharp{H}_\text{c}$ is expressed as
\begin{equation}
\mathrm{SNR}_{\mathrm{avg}}=\frac{1}{N_\text{c}} \sum_{i=1}^{N_\text{c}} \mathrm{SNR}_i,
\end{equation}
% \begin{equation}
% \label{snrequ}
% \text{SNR}=\frac{\sum_{n=1}^{N_\text{c}}\left\|h_n \cdot s_n\right\|^2}{\sum_{n=1}^{N_\text{c}} \sigma_{n}^2},
% \end{equation}
where $\mathrm{SNR}_{\text {avg }}$ is the average SNR across all subchannels, $N_\text{c}$ is the total number of subchannels, and $\mathrm{SNR}_i$ is the SNR for the $i$-th subchannel. Therefore, the received data of the $i$-th subchannel at the receiver side ${s^{'}_i}$ can be expressed as
% where, $\sigma_{n}^{2}$ represents the noise within the $n$th subchannel. The received bits after the wireless channel, marked as ${s^{'}_n}$, are articulated by the subsequent equation, which can be expressed as
\begin{equation}
{s^{'}_i}=s_i \otimes h_i + \textcolor{black}{\omega_{i}},
\end{equation}
\textcolor{black}{where the symbol $\otimes$ refers to circular convolution, an operation correlating the input signal with a finite impulse response, and $\omega_{i}$ represents the noise in the $i$-th subchannel. The channel response value $h_i$, varies due to frequency-selective fading in the FDM system. Each subchannel response $h_i$ is assumed to be a complex Gaussian random variable. Subsequently, the received data, denoted as ${s^{'}_i}$, is processed by both a traditional channel decoder and a source decoder at the receiver to recover the original data.}
\subsection {Novel Goal-oriented Semantic communication Framework}
\begin{figure*}[ht]
  \centering
  \includegraphics[width=\textwidth]{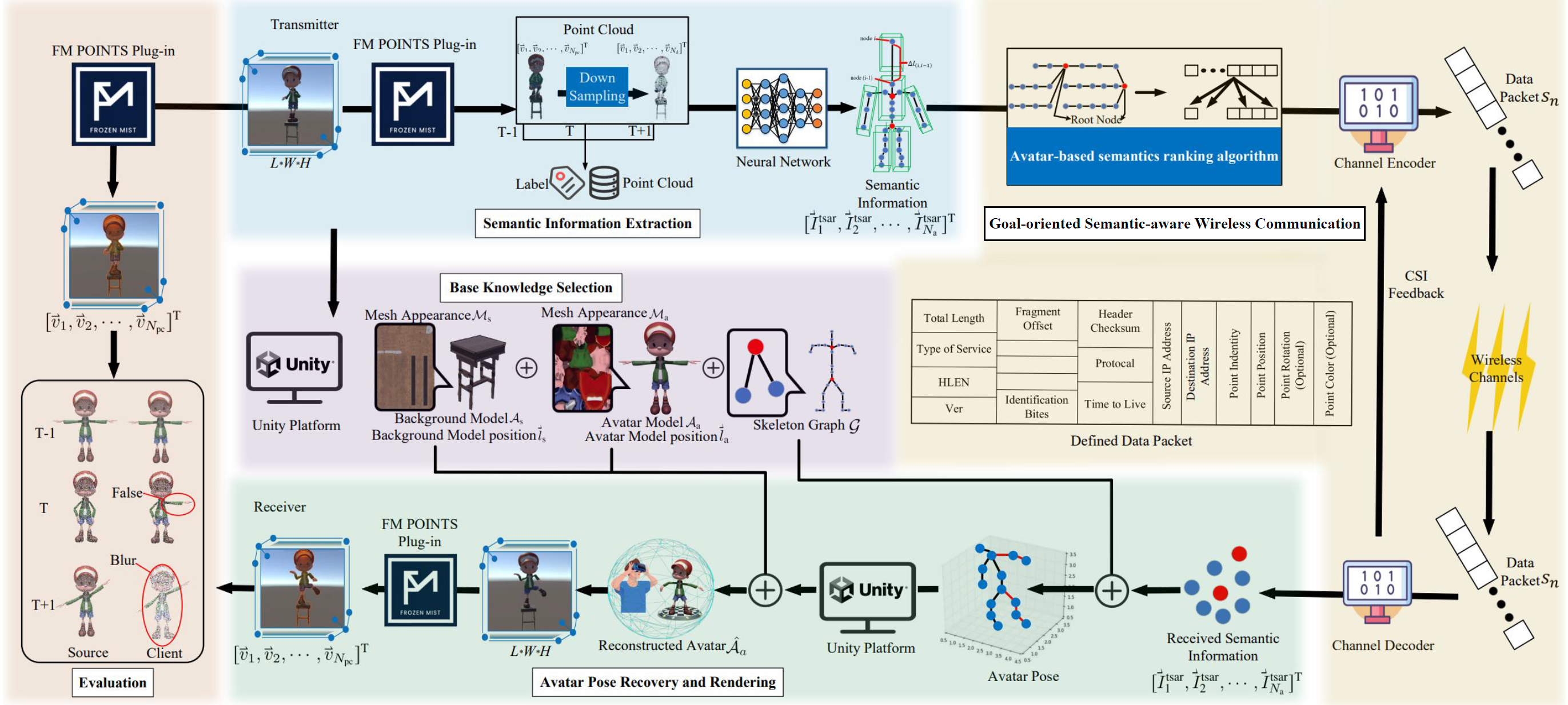} %1.png是图片文件的相对路径
  \caption{Goal-oriented semantic communication framework} %caption是图片的标题
  \label{semanticsframework} %此处的label相当于一个图片的专属标志，目的是方便上下文的引用
\end{figure*}

\textcolor{black}{In this section, we provide a detailed description of our proposed GSAR framework, that not only compare with the traditional point cloud communication framework but also incorporates several goal-oriented strategies, including effectiveness level optimization methodology.}
% In this section, we provide a detailed description of our proposed TSAR framework to \textcolor{black}{highlight the significant differences from the traditional point cloud communication framework.} 
The GSAR framework leverages shared base knowledge and utilizes a goal-oriented context at the semantic level, to exploit more efficient and effective communication for AR application.
As illustrated in Fig. \ref{semanticsframework} in the next page, the modules in GSAR include semantic information extraction, goal-oriented semantic wireless communications, avatar pose recovery and rendering. 
% These components function together to provide a cohesive, task-oriented, and semantics-aware AR experience.
% The TSAR procedure diverges substantially from those used in traditional point cloud communication frameworks, possibly leveraging shared base knowledge and prioritizing the analysis of the more task-relevant context at the semantics level. As depicted in Figure \ref{semanticsframework}, the stages encompassed in TSAR consist of semantics information data sensing and processing, avatar-based semantics ranking algorithm, wireless communication, human body position reconstruction, and AR rendering and display.

% In this section, we focus on analyzing the characteristics of all data types in the TASO framework, summarizing the system model with corresponding extraction methods, and proposing the objective function, as shown in Fig. \ref{Fig_1}. 
% \subsubsection {Architecture Detail}
% In the 3D scene at time slot $t$, different objects $O_i$, such as tables and avatars, are denoted as $O_i=\{O_1,..., O_k\}$, where $k$ represents the number of models in the scene. 

\subsubsection {Semantic Information Extraction}
\textcolor{black}{Unlike traditional point cloud communication framework, which primarily relies on raw point cloud data for AR scenery representation and transmission, our proposed GSAR framework provides a more sophisticated approach to extract a rich depth of semantic and effectiveness levels data from the raw point cloud.
% This approach enables simultaneous processing and compression of raw point cloud data at both task and semantics levels, thus provide a more efficient framework for data management in AR applications.
}
% This significantly reduces data size, as only semantics and task-related data are extracted and transmitted within the TSAR system, enhancing its efficiency.
The process begins with the downsampled point cloud sensing data, $\mathbf{P}_\text{dpc}$, as the input. This point cloud data encapsulates all the AR scenery, which are broadly divided into two categories: the moving avatar model $\mathcal{A}_\text{a}$ and the stationary model $\mathcal{A}_\text{s}$. Only the avatar's moving position is considered essential information and needs to be refreshed at every frame. 
% \textcolor{black}{Thus the output of this semantics information extraction process is the skeletons information of the moving avatar, which can be represented as \overrightharp{I}, each skeleton information includes both the three-dimensional location $l^{\text{tsar}}_i=(l_\text{x}, l_\text{y}, l_\text{z})$ and quaternion rotation $r^{\text{tsar}}_i=(r_\text{x}, r_\text{y}, r_\text{z}, r_\text{w})$.}
Thus, the output of this semantic information extraction process is the skeletons information of the moving avatar, $\overrightharp{I}^\text{gsar}_i$, which can be represented as 
\begin{equation}
\label{semanticinfor}
\overrightharp{I}^\text{gsar}_i=(\overrightharp{l}_i, \overrightharp{r}_i)=(l_\text{x}, l_\text{y}, l_\text{z}, r_\text{x}, r_\text{y}, r_\text{z}, r_\text{w}), \ i \in [0,{N}_{\text{a}}],
\end{equation}
where $N_\text{a}$ represents the total number of skeletons in the avatar, $\overrightharp{l}_i$ represents the three-dimensional location and  $\overrightharp{r}_i$ represents the quaternion rotation of the $i$-th skeleton in the avatar model.
% Different from on the data sensing and acquisition of the traditional point cloud communication, our proposed TSAR framework advocates for simultaneous point cloud data processing and compression within both task and semantics levels, utilizing deep learning algorithms to significantly reduce data size. Only semantics and task-related data is extracted and transmitted in the TSAR system.
% Specifically, the models in AR scenery are grouped as moving avatar model $A$ and stationary model $S$. Within the AR scenery, only the moving position of an avatar is deemed essential information. Each avatar model comprises a varying number of skeleton points, like wrists, elbows, and knees, and each skeleton contains information location vector ($l_\text{x}, l_\text{y}, l_\text{z}$) and quaternion rotation ($r_\text{x}, r_\text{y}, r_\text{z}, r_\text{w}$). 

Apart from quaternion rotation, current research also employs euler angles to represent rotations in AR scenery. In comparison to quaternion, euler angles offer a simpler and more information-efficient method to represent rotation and calculate root node position when a fixed root node point is available. This approach needs less information to reconstruct the avatar's pose compared to quaternion, resulting in less data packets and potentially more efficient communication \cite{quintero2022excite}. The transformation from rotation to euler angles can be expressed as
\begin{equation}
\label{eulerangle}
{ 
\left[\begin{array}{c}
{e_\text{p}} \\
{e_\text{r}} \\
{e_\text{y}}
\end{array}\right]=\left[\begin{array}{c}
\arctan \large \frac{2\left(r_\text{y} r_\text{z}+r_\text{w} r_\text{x}\right)}{1-2\left(r_\text{x}^2+r_\text{y}^2\right)} \\
\arcsin \left(2\left(r_\text{w} r_\text{y}-r_\text{x} r_\text{z}\right)\right) \\
\arctan \frac{2\left(r_\text{x} r_\text{y}+r_\text{w} r_\text{z}\right)}{1-2\left(r_\text{y}^2+r_\text{z}^2\right)}
\end{array}\right] *\frac{180}{\pi}.
}
\end{equation}
where the $e_\text{p}$, $e_\text{y}$, and $e_\text{r}$ are defined as the pitch, roll, and yaw in euler angles to represent rotations around the three primary axes with an associated root point. 
The semantic information of the AR application, denoted as $\mathbf{D}_{\text{gsar}}$, represents all the skeleton information $\overrightharp{I}^{\text{gsar}}_i$ of the avatar model generated through a semantic information extraction process from the downsampled point cloud $\mathbf{P}_{\text{dpc}}$, which can be expressed as
\begin{equation}
\mathbf{D}_\text{gsar} = [{\overrightharp{I}^{\text{gsar}}_\text{1}, \overrightharp{I}^{\text{gsar}}_\text{2}, \cdots, \overrightharp{I}^{\text{gsar}}_{{N}_{\text{a}}}}]^{\text{T}} = \mathcal{S}(\mathbf{P}_\text{dpc}, {\theta}_\text{s}) ,
\end{equation}
where $\mathcal{S}(\cdot)$ represents the semantic information extraction process, and ${\theta}_\text{s}$ encompasses all the experimental and neural network parameters.
This equation represents the entire semantic information extraction process, which maps the downsampled point cloud data $\mathbf{P}_{\text{dpc}}$ to a more meaningful semantic representation $\mathbf{D}_{\text{gsar}}$ for further transmitting over wireless channels.

\subsubsection {Goal-oriented Semantic Communications}
Building upon the extracted semantic information, we develop an avatar-based semantic ranking algorithm to integrate goal-oriented semantic information ranking into end-to-end wireless communication
to exploit the importance of semantic information to an avatar-based AR displaying application. 
The algorithm correlates the importance evaluation of semantic information and goal relevance with channel state information feedback, thereby prioritizing more important semantic information for optimal transmission over more reliable subchannels.
More specifically, each skeleton is represented as a node in the avatar skeleton graph $\mathcal{G}$ as shown in the Fig. \ref{rankingFigprocesss}, \textcolor{black}{and the skeleton ranking is determined by a calculated weight in the skeleton graph, which indicates the level of importance in the later avatar pose recovery.} The weights of all semantic information $\mathbf{D}_\text{gsar}$ are denoted as $\overrightharp{W}_\text{gsar}$ and can be formulated as
% Based on the extracted semantics information, TSAR utilizes an avatar-based semantics Ranking Algorithm to incorporate task-level optimization, further improving avatar transmission performance in AR applications. This approach evaluates the essentiality and importance of all semantics information in relation to one another. To be more specific, the structure of the avatar is formulated as a Structure Graph (SG), where each skeleton represents a node. The Euler distance between two skeleton points is represented as $d$. Consequently, the weights of the avatar's skeleton information, denoted as $W_{I^{A_i}}$, can be expressed as:
\begin{equation}
\overrightharp{W}_{{\text{gsar}}} =[{\omega}_{I_1},{\omega}_{I_2} ..., {\omega}_{I_{N_\text{a}}}]^\text{T}= \mathcal{W}(\mathbf{D}_\text{gsar}, \mathcal{G}),
\end{equation}
where ${w}_{I_i}$ represents the weight of the semantic information of the $i$-th skeleton in avatar skeleton graph, these node weights essentially represent the importance of the semantic information to the avatar representation, with higher weights indicating greater importance of the skeleton information for avatar pose recovery. By correlating these weights representing the importance of semantic information with CSI feedback during wireless communication, the effectiveness of the avatar transmission in AR application could be optimized. Specifically, the semantically important information is mapped and transmitted over more reliable subchannels.
Current research in the FDM has demonstrated that CSI can be accurately estimated at the transmitter side using suitable algorithms and feedback mechanisms \cite{thoota2022massive}.
% This approach builds upon existing research in OFDM, which has shown that 
Consequently, the subchannel gains $h_i$ at the receiver side are assumed to be added in the CSI feedback, enabling the transmitter to be aware of the accurate all the subchannel state in the FDM.
According to Eq. (\ref{snrequ}), the subchannel with a higher SNR will have a better subchannel state and thus achieve a more reliable transmission for semantic information. 
Therefore, an ascending sorting is employed to establish a mapping function $\mathcal{M}(\cdot)$ between the semantic information and various subchannels. This mapping relies on the weights calculated for the semantic information and the CSI. Higher weights, indicating greater importance of the semantic information in the avatar pose recovery, are assigned to more reliable subchannels. The mapping function is expressed as
% This strategy ensures enhanced protection and transmission of the avatar's semantics information.
% Thus, an ascending sorting algorithm is implemented to map semantics information with different subchannels based on the calculated weights and CSI for better avatar semantics information protection. The channel mapping process $M(\cdot)$ is expressed as follows:
\begin{equation}
\label{equationmap}
\begin{array}{r}
\mathcal{M}(\overrightharp{W}_{\text{gsar}},\mathcal{G}, \overrightharp{H}_\text{c})
=\{\overrightharp{I}^{\text{gsar}}_{i}, h_{j}\} ,i \in[1, N_\text{a}], j \in[1, N_\text{c}],
\end{array}
\end{equation}
\textcolor{black}{the map $\{\overrightharp{I}^{\text{gsar}}_{i}, h_{j}\}$ refers to transmitting the semantic information $\overrightharp{I}^{\text{gsar}}_{i}$ on the subchannel $h_{j}$. Based on this channel mapping, semantic information with higher priority is being mapped to subchannels with better channel responses.}
% where the map $\{\overrightharp{I}^{\text{tsar}}_{i}, h_{j}\}$ refers to transmit the semantic information $\overrightharp{I}^{\text{tsar}}_{i}$ at the subchannel $h_{j}$.
% Based on the channel mapping results, each semantic information is transmitted through different subchannels in the OFDM subchannels. 
% The received data of TSAR can be denoted as follows:
% \begin{equation}
% \mathbf{D}^{'}_{\text{tsar}}= \{I^{'}_1, I^{'}_2,...I^{'}_{N_{a}}\},
% \end{equation}
% where $\mathbf{D}^{'}_{\text{tsar}}$ and $I^{'}_i$ represent all the received semantics information and single received skeleton information, respectively. By doing so, we can optimize the transmission of semantics information over the wireless physical channel based on subchannel quality, ensuring efficient and accurate delivery of more important semantics information. 
% This method allows for better adaptation to various communication scenarios and helps enhance the overall performance of AR applications in terms of avatar transmission and viewing experience for clients.
% The details of the weights calculation AbSR algorithm $W(\cdot)$ will be described in the following paragraph.

% The transmitted signal after passing through the physical channel is represented as:

% where $\theta$ denotes the parameters involved in channel encoding.

% \subsubsection{Base Knowledge Update}
% 
\subsubsection {Avatar Pose Recovery and Rendering}
In contrast to traditional point cloud wireless communication framework, the GSAR framework approaches avatar pose recovery differently with the transmission of the base knowledge at the beginning of AR application. As illustrated in Fig. \ref{semanticsframework}, the data could be used for base knowledge $\boldsymbol{B}_{\text{*}}$
% the base knowledge of TSAR framework $\boldsymbol{B}_{\text{tsar}}$ 
encompasses different types of information, which include avatar skeleton graph $\mathcal{G}$, avatar initial position $l_o$, avatar model $\mathcal{A}_\text{a}$, stationary background model $\mathcal{A}_\text{s}$, stationary initial position $l_s$, and their respective appearance meshes, $\mathcal{M}_\text{a}$ and $\mathcal{M}_\text{s}$. Whenever a new 3D object appears in the AR scenery, the base knowledge at both transmitter and receiver need to be updated synchronously. 

In this way, the GSAR framework considers the avatar as a whole entity and recover the avatar's pose using a limited set of skeleton points instead of treating individual points as the smallest recovery unit.
% Instead of treating individual points as the smallest recovery unit, the TSAR framework considers the avatar as a whole entity. 
% This approach is possible due to the base knowledge shared at the beginning of the AR application, which includes the avatar model.
% This base knowledge allows the client to recover the avatar's pose using a limited set of skeleton points, instead of processing extensive point cloud data. Thus, the TSAR framework prioritizes the transmission of crucial avatar skeleton points, enabling a more efficient and streamlined recovery of the avatar's pose. This approach not only reduces the amount of data needed for transmission but also simplifies the task of avatar pose recovery at the client side.
% Compared to traditional point cloud wireless communication frameworks, another significant difference in the TSAR framework lies in the avatar pose recovery module, which treats the avatar as a complete object instead of using the single point as the smallest recovery unit. 
% The TSAR framework leverages shared base knowledge at the onset of the AR application, requiring the transmission of avatar skeleton points rather than a multitude of points for avatar pose recovery. 
% The shared base knowledge comprises the .
% This approach requires significantly less data, resulting in more efficient avatar recovery. 
The avatar pose recovery process $\mathcal{R(\cdot)}$ can be expressed as
% A significant difference between TSAR and traditional point cloud frameworks is that TSAR incorporates an avatar position reconstruction process, treating the avatar and other background models as a whole object rather than using point clouds as the minimum recovery unit. With considerably less information compared to numerous point clouds, TSAR human position recovery is carried out using shared base knowledge at the first frame, such as the avatar appearance mesh $M_{A_i}$ and the skeleton connection graph $SG$. The skeleton in the avatar model are typically represented as a tree-based pictorial structure. More specifically, the avatar reconstruction process $R(.)$ can be expressed as:
\begin{equation}
\label{recoverpos}
\hat{\mathcal{A}}_a=\mathcal{R}(\mathbf{D}^{’}_\text{gsar}, \boldsymbol{B}_\text{gsar}),
\end{equation}
where $\boldsymbol{B}_{\text{gsar}}$ represents the base knowledge of GSAR, and $\hat{\mathcal{A}}_a$ denotes the avatar model ${\mathcal{A}_\text{a}}$ with appearance $\mathcal{M}_\text{a}$ after pose recovery with semantic information $\mathbf{D}^{'}_\text{gsar}$.

The AR displaying process is quite straightforward by presenting the reconstructed avatar $\hat{\mathcal{A}}_\text{a}$ and the stationary background model $S_o$ in the AR scenery. 
% The process involves associating each skeleton information $I$ with the avatar model $A_o$ in the Unity3D platform, eliminating the need to refresh the entire point cloud for every frame. This approach streamlines the rendering process, thereby improving the QoS in the AR application.
The process of avatar pose recovery in the GSAR framework is intricately designed and hinges on associating each piece of skeleton information $\overrightharp{I}^{\text{gsar}}_i$ with the avatar model $\mathcal{A}_a$ on the Unity3D platform. 
% This association step is critical as it allows for a significant reduction in the data required for each frame rendering.
In traditional point cloud communication frameworks, the entire point cloud data must be refreshed for each frame, which can be a computationally expensive and time-consuming process. In contrast, the GSAR framework only requires the updating of the skeleton information associated with the avatar's movements, and update the avatar's pose based on these information. 
% This dramatically decreases the amount of data that needs to be processed and transmitted for each frame.
% Once the skeleton points have been updated, they are used to adjust the avatar's pose on the Unity3D platform. This is done by modifying the associated points on the avatar model $A_o$. As a result, the avatar's movements are rendered smoothly and accurately, mirroring the movements captured by the semantics sensing process.
% This streamlined approach not only reduces computational requirements but also improves the Quality of Service (QoS) in the AR application. By minimizing data processing and transmission, it allows for faster rendering and less latency, leading to a more immersive and responsive user experience. Users are thus provided with an AR experience that is as seamless and realistic as possible, enhancing their engagement with the AR environment.

% The whole process of AR video frame playing delay time is denoted as $T_p$.
% As for the rendering and playing process of TOSAAR, with the baskknowlege of the whole avatar appareance mesh $M_i$, the Unity3D platform only need to set the skeleton positoon $S_i$ with the avatar model, but not need to refresh all the point cloud every frame, the whole process of AR video frame playing delay time is represents as $T_p$.

\subsection {Problem Formation}
In summary, the overall framework aims to achieve goal-oriented semantic communications with efficient data transmission for better avatar representation in wireless AR applications. The primary objective of the framework is to 
% minimize the transmitted data packets in wireless communication and 
maximize the client-side AR viewing experience based on the transmitted semantic information. The objective function can be represented as
% \begin{equation}
% \mathcal{P}: \min _{\theta_S,\left(I_i, h_j\right)} \lim _{T \rightarrow+\infty} \frac{1}{T} \sum_{t=0}^T \sum_{i=0}^{N_A}\left(I_i^t-\hat{I}_i^t\right) \cdot W_{I_i}
% \end{equation}
\begin{equation}
\begin{aligned}
& \mathcal{P}: \min _{\{\theta_\text{s},\left(\overrightharp{I}_i, h_j\right)\}} \lim _{T \rightarrow+\infty} \frac{1}{T} \sum_{t=0}^T \sum_{i=0}^{N_a}\left(\overrightharp{I}_{i,t}^{\text{gsar}}-\overrightharp{I}_{i,t}^{{\text{gsar}'}}\right) \cdot \omega_{I_i}, \\
& \text { s.t. } \ \ \ \  i \in[1, N_\text{a}], \ \ j \in[1, N_\text{c}],
\end{aligned}
\end{equation}
where $\overrightharp{I}_{i,t}^{\text{gsar}}$ represents the semantic information of the $i$-th skeleton at time $t$, and $\overrightharp{I}_{i,t}^{\text{gsar}'}$ is the received semantic information after the wireless channel. The weights $\omega_{I_i}$ reflect the importance of each skeleton node $i$ in representing the avatar graph. This equation formulates the problem of minimizing the error in avatar representation during transmission.

\begin{table*}[t]
\centering
\caption{\textcolor{black}{Transmit Message}}
\label{semanticinfotable}
\begin{tabular}{|l|c|c|}
\hline
\textcolor{black}{Framework} & \textcolor{black}{Semantic Information} & \textcolor{black}{Base Knowledge} \\ \hline
\textcolor{black}{GSAR}      & $\overrightharp{I}^{\text{gsar}}$=$(l_\text{x}, l_\text{y}, l_\text{z}, r_\text{x}, r_\text{y}, r_\text{z}, r_\text{w})$                    & $\boldsymbol{B}_{\text{gsar}}$=$\{\mathcal{A}_\text{o}, \mathcal{A}_\text{s}, \mathcal{M}_\text{o}, \mathcal{M}_\text{s}, \overrightharp{l}_\text{s} \}$                \\ \hline
\textcolor{black}{E-GSAR}    & $\overrightharp{I}^{\text{egsar}}$=${(e_\text{r}, e_\text{y}, e_\text{p})}$                   & $\boldsymbol{B}_{\text{egsar}}$=$\{\mathcal{M}_\text{a}, \mathcal{M}_\text{s}, \mathcal{A}_\text{a}, \mathcal{A}_\text{s},\overrightharp{l}_\text{a}, \overrightharp{l}_\text{s}, \mathcal{G}\}$                \\ \hline
\textcolor{black}{EC-GSAR}   & $\overrightharp{I}^{\text{ecgsar}}$=${(e_\text{r}, e_\text{y}, e_\text{p})}$                    & $\boldsymbol{B}_{\text{ecgsar}}$=$\{\mathcal{M}_\text{a}, \mathcal{M}_\text{s}, \mathcal{A}_\text{a}, \mathcal{A}_\text{s},\overrightharp{l}_\text{a}, \overrightharp{l}_\text{s}, \mathcal{G}\}$                \\ \hline
\end{tabular}
\end{table*}

\section {Semantic Level Design}
% In this section, we will elaborate on the modules within the TSAR framework, which incorporates the semantics-level content of the AR application and optimizes the processes of avatar transmission and recovery and present the evaluation metrics for assessing the impact of these modules on performance in the wireless semantics communication of the TSAR framework. The modules involved in semantics level optimization include deep learning-based semantics sensing, human pose recovery, and AR playing.
In this section, we will discuss the semantic extraction and recovery blocks, including semantic information extraction with deep learning, base knowledge selection, avatar pose recovery, and evaluation metric.
% In this section, we will detail the process of TSAR, which incorporates the semantics-level content of the AR application and how to optimize the problem. The semantics communication modules include semantics sensing with deep learning, human pose recovery, and AR video playback. We then provide the evaluation matrix for these modules' impact on performance in the semantics communication of TSAR.

\subsection{Semantic Extraction with Deep Learning}

\begin{figure}[t]
\centering
% \label{networkarchitecture}
\includegraphics[width=8.5cm]{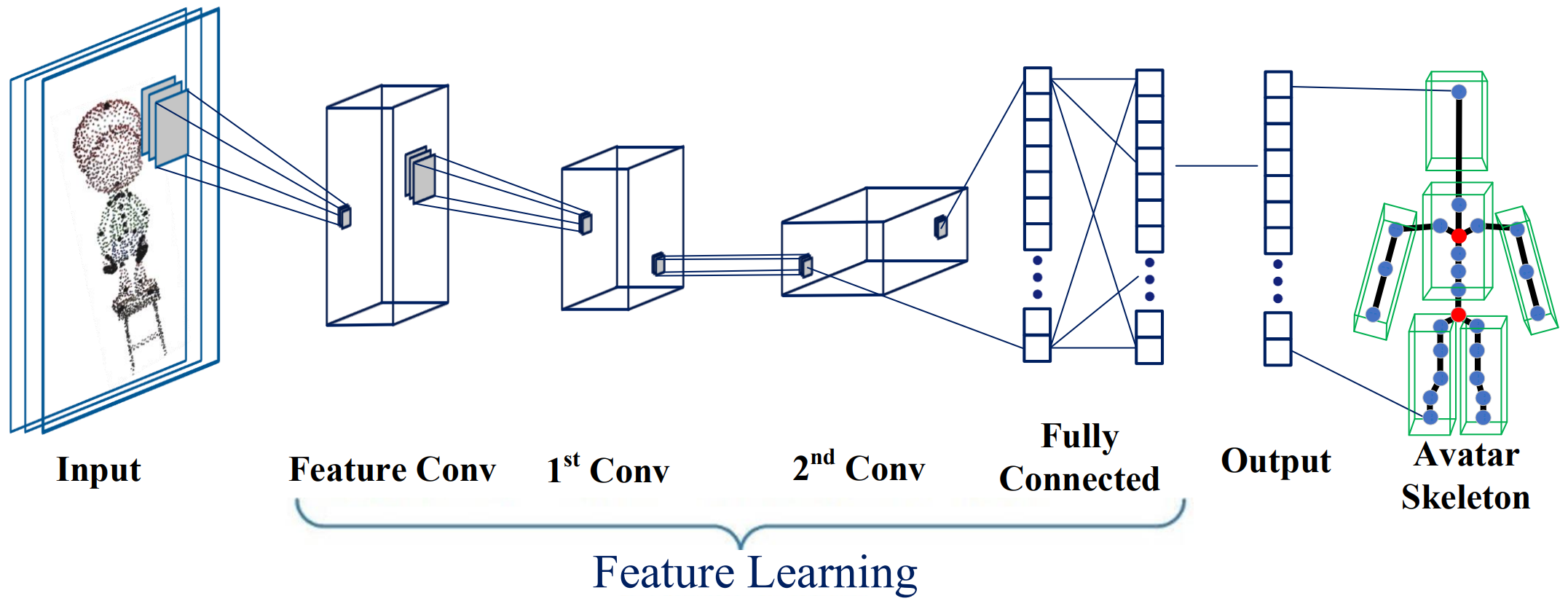}
% [height=4.5cm]表示高度
%[width=9.5cm]表示宽度
%{111.eps}表示eps格式的图片，名为111
\caption{Semantic information extraction network}
%图片的名称
\label{rankingFigprocesss1}
%图片的标签，用于文章中的引用，注意到标签的数字与实际文章显示的数字可能不同
\end{figure}

Inspired by the KeypointNet proposed in \cite{you2020keypointnet}, we propose a semantic extraction network called SANet to extract the skeleton keypoint information of a moving avatar from the whole point cloud of AR scenery. The extraction is an integral step towards creating a more interactive and immersive augmented reality experience.
The SANet operates by using downsampled point cloud data ${\mathbf{P}_{\text{dpc}}}$ as input, which represents the 3D coordinates of both the stationary models and the moving avatar. This data is then processed by the SANet to extract accurate avatar skeleton information, crucial for reproducing the avatar's movements in the virtual environment.
The design objective of the SANet is to minimize the Euclidean distance ($\mathcal{L}_2$) between the predicted semantic information, denoted as $\mathcal{S}({\mathbf{D}_{\text{dpc}}})$, and the labeled semantic information of the skeleton location, represented as $\mathbf{D}^{l}_{\text{gsar}}$. The interplay between these variables is captured as
\begin{equation} \text{Loss}=\arg \min \underset{\left(\theta_{\text{s}}\right.)}{\mathcal{L}_2}\left(\mathcal{S}({\mathbf{P}_\text{dpc}}), \mathbf{D}^{l}_{\text{gsar}}\right). \label{Losstheta} \end{equation}
where $\theta_{\text{s}}$ represents all the neural networks and experiment parameters in the SANet, which is defined in Table \ref{tab:Experiment_Parameter} and Fig. \ref{rankingFigprocesss1}. Training the SANet involves optimizing these parameters to minimize the loss, thus enhancing the accuracy of semantic information extraction.

To determine the most suitable backbone for the designed SANet, we train the SANet with various backbone networks, including ResNet, RsCNN, PointNet, SpiderCNN, PointConv, and DGCNN \cite{qiu2021dense}. Similar to \cite{you2020keypointnet}, we use the mean average precision (mAP) as the performance evaluation metric to assess the semantic information extraction accuracy of the predicted keypoint probabilities in relation to the ground truth semantic information labels.
% To establish the most suitable backbone design and outstanding performance for our SANet, we train the SANet with various backbone networks, including ResNet, RS-CNN, PointNet, SpiderCNN, PointConv, and DGCNN\cite{qiu2021dense}. As in the paper \cite{you2020keypointnet}, the mean Average Precision (mAP) is employed to assess the prediction accuracy of the predicted keypoint probabilities relative to the ground truth labels. Fig. \ref{Fig_2Training_results} demonstrates that the SpiderCNN-based SANet achieves the highest accuracy in skeleton information extraction, exceeding 96\% among the same training epochs. This result renders the SANet significantly more reliable and stable compared to other backbone-based networks.

% \subsection{}

\begin{table}[t]
  \centering
  \caption{SANet parameters and training setup}
  \label{tab:Experiment_Parameter}
  \begin{tabular}[l]{@{}lc}
  \toprule
      \textbf{Parameter} & \textbf{Value} \\
      \midrule
      \footnotesize \textbf{\emph{Cell}} \\ 
      \footnotesize Semantic network & \footnotesize In (2048,3), out (25,1) \\
      \footnotesize Feature conv &  \footnotesize(In feature=2048, out feature=1440) \\ 
      \footnotesize $1^\text{st}$ Conv2d & \footnotesize(In feature=256, out feature=256) \\ 
      \footnotesize $2^\text{nd}$ Conv2d & \footnotesize(In feature=256, out feature=128) \\ 
      % \footnotesize spider conv4 & \footnotesize(In feature=128, out feature=25) \\ 
      \footnotesize Output layer & \footnotesize (In feature=128, out feature=25) \\ 
      \midrule
      \footnotesize \textbf{\emph{Simulation}} \\
      \footnotesize \textcolor{black}{Learning rate} & \footnotesize  \textcolor{black}{$10^{-4}$} \\ 
      \footnotesize \textcolor{black}{Optimizer} & \footnotesize \textcolor{black}{Adam} \\ 
      \footnotesize \textcolor{black}{Episode} & \footnotesize \textcolor{black}{900} \\ 
      \footnotesize \textcolor{black}{Batch size} & \footnotesize \textcolor{black}{16} \\ 
      \footnotesize \textcolor{black}{Loss Function} & \footnotesize \textcolor{black}{MSE} \\ 
      \footnotesize \textcolor{black}{Momentum} & \footnotesize \textcolor{black}{SGD} \\ 
      \footnotesize \textcolor{black}{Activation function} & \footnotesize \textcolor{black}{ReLU} \\ 
    \bottomrule
  \end{tabular}
\end{table}

\subsection{Base Knowledge Selection}
% As illustrated in Fig. \ref{semanticsframework}, the TSAR framework encompasses different types of data: the avatar skeleton graph, moving avatar model, stationary background model, and their corresponding appearance mesh. Whenever a new object appears in the AR scene, the base knowledge at the transmitter needs to be updated synchronously with the receiver with the new model. 
% To better explore the most suitable base knowledge, we have designed the following ablation experiments for semantic communication with different shared base knowledge and semantic information definitions\footnote{Semantic information, as shown in Fig. \ref{semanticsframework}, consists of the skeleton information that need to be transmitted in every frame. Conversely, base knowledge encompasses information used primarily in the first frame.}, which include the base TSAR framework (TSAR) and euler angle based TSAR framework (E-TSAR).
To better explore the most suitable base knowledge, we propose basic GSAR framework (GSAR) and euler angle based GSAR framework (E-GSAR) that considers different shared base knowledge and semantic information definition\footnote{Semantic information, as presented in Table \ref{semanticinfotable}, consists of the skeleton information that need to be transmitted in every frame. Conversely, base knowledge encompasses information used primarily in the first frame.}. 

\textbf{GSAR}: 
% Inspired by \cite{liu2022deep}, the authors employ quaternion rotation and vector position, along with the avatar model, to restore the avatar's position. 
For the basic GSAR framework, semantic information for each skeleton is defined as the data pertaining to position and quaternion rotation as in Eq. (\ref{semanticinfor}). The shared base knowledge, denoted as $\boldsymbol{B}_{\text{gsar}}$, comprises the stationary background model, stationary model initial position moving avatar model, and their corresponding appearance meshes, which is denoted as
\begin{equation}
\boldsymbol{B}_{\text{gsar}}=\{\mathcal{A}_\text{o}, \mathcal{A}_\text{s}, \mathcal{M}_\text{o}, \mathcal{M}_\text{s}, \overrightharp{l}_\text{s} \}.
\end{equation}

\textbf{E-GSAR}: 
% Inspired by \cite{bernardes2022quaternion}, the authors utilize vector euler angles to represent quaternion for skeleton rotation. 
As an extension of GSAR, the semantic information in each skeleton $I_i$ is defined as the euler angle rotation in E-GSAR, according to Eq. (\ref{eulerangle}), which could be defined as
\begin{equation}
\label{semanticsinforationetasr}
\overrightharp{I}^{\text{egsar}}_i =(\overrightharp{e}_i)= {(e_\text{r}, e_\text{y}, e_\text{p})}, \ i \in [0,{N}_{\text{a}}],
\end{equation}
where the shared base knowledge $\boldsymbol{B}_{\text{egsar}}$ encompasses the avatar skeleton graph, avatar initial position, stationary background model, stationary model initial position,  moving avatar model, and their appearance meshes, defined as
\begin{equation}
\label{baseknowledgedaiG}
\boldsymbol{B}_{\text{egsar}}=\{\mathcal{M}_\text{a}, \mathcal{M}_\text{s}, \mathcal{A}_\text{a}, \mathcal{A}_\text{s},\overrightharp{l}_\text{a}, \overrightharp{l}_\text{s}, \mathcal{G}\}.
\end{equation}

\subsection{Avatar Pose Recovery}
% The details of the human pose recovery involve using the SG in the base knowledge and the received semantics information to reconstruct the avatar position, significantly reducing the recovery time. Specifically, a recursive algorithm is employed to traverse and assign all skeleton information and append values to the skeletons. However, due to differences in the definition of the semantics information and the shared base knowledge, the human pose recovery process has slight variations between the base TSAR and E-TSAR framework.
The avatar pose recovery involves using the skeleton graph $\mathcal{G}$ in the base knowledge and the received semantic information to reconstruct the avatar pose. The entire avatar pose recovery process is shown in $\textbf{Algorithm \ref{humanposerecovery}}$. Specifically, a recursive algorithm is employed to traverse and assign all skeleton information to the avatar model $\mathcal{A}_a$ with initialized parameters. However, due to differences in the definition of the semantic information and the shared base knowledge, the avatar poses recovery process has variations between the GSAR and E-GSAR framework.

% Compared to the traditional point cloud communication framework, human pose recovery is utilized to reconstruct the avatar position using the base knowledge (SG) and the received semantics information, significantly reducing recovery time. The design principle aims to recover the avatar's body shape and pose information extracted from the transmitted semantics information, $I_i$. Specifically, a recursive algorithm is employed to traverse and assign all skeleton information and append values to the skeletons. The detailed implementation process is illustrated in Algorithm \ref{}. The human pose recovery process has slight differences between the TSAR and E-TSAR methods, depending on the defined base knowledge and semantics information.

% On one hand, TSAR applies a straightforward recovery method, assigning the skeleton information directly according to the skeleton name ID with the received position and rotation quaternion information. On the other hand, with only the Euler angle of each skeleton information transmitted, E-TSAR first calculates the skeleton position according to the skeleton graph (SG). To be more specific, the location of skeleton node $l_{i}$ is represented as:
On the one hand, the basic GSAR framework employs a simple avatar pose recovery method, assigning the avatar model with value based on the skeleton point identity using the received position vector and quaternion rotation. On the other hand, the E-GSAR framework, which only transmits the euler angle of each skeleton point as semantic information, requires calculating each skeleton position with respect to its root point in the skeleton graph before assigning the skeleton information to the avatar model. The E-GSAR framework reconstructs the avatar pose by first determining the relationships between the skeleton points in the avatar skeleton graph $\mathcal{G}$. It then computes the position of each skeleton point by considering its euler angle and the position of its root point within the $\mathcal{G}$, the relative distance vector $\Delta \overrightharp{l}_{(i,i-1)}$ between the $i$-th skeleton node and the previous ${(i-1)}$-th node can be represented as
\begin{equation}
\label{eq_l11}
\Delta \overrightharp{l}_{(i,i-1)} =(\Delta\text{x},\Delta\text{y},\Delta\text{z})= \overrightharp{e}_i \times \overrightharp{l}_{i-1},
\end{equation}
where $ \overrightharp{e}_i$ represents the eular angle of the $i$-th skeleton node, $(\Delta\text{x},\Delta\text{y},\Delta\text{z})$ represents the distance between two skeleton node towards the x, y, and z coordinates, and the actual position of the $i$-th skeleton node will be calculated by combining $\Delta \overrightharp{l}_{(i,i-1)}$ and $\overrightharp{l}_{i-1}$, which can be expressed as
\begin{equation}
\label{eq_l12}
\overrightharp{l}_i= \overrightharp{l}_{i-1} + \Delta \overrightharp{l}_{(i,i-1)},
\end{equation}
where the root node position $\overrightharp{l}_0$ is equal to the avatar initial position $\overrightharp{l}_\text{a}$ in the base knowledge, and $\overrightharp{l}_i$ represents the position of the $i$-th skeleton node in the avatar,
with its three components representing the x, y, and z coordinates respectively.
% \STATE (Avatar pose recovery for the TSAR)
\begin{algorithm}[t]
	%\textsl{}\setstretch{1.8}
	\renewcommand{\algorithmicrequire}{\textbf{Input:}}
	\renewcommand{\algorithmicensure}{\textbf{Output:}}
	\caption{Avatar Pose Recovery}
        \label{humanposerecovery}
	\begin{algorithmic}[1]
		\STATE Initialization: Received base knowledge $\boldsymbol{B}_{\text{*}}$, received data $\mathbf{D}^{'}_{\text{gsar}}$
            \STATE Get skeleton graph $\mathcal{G}$, avatar initial position $\overrightharp{l}_a$ avatar model $\mathcal{M}_a$, and avatar appearance mesh $\mathcal{A}_a$ from $\boldsymbol{B}_{\text{*}}$
		\STATE  Count the skeleton number $N_\text{a} = \mathbf{C}_\text{s}(\mathcal{G}) $
  		\STATE  Count the received semantic information $N_\text{r} = \mathbf{C}_\text{r}(\mathbf{D}^{'}_{\text{gsar}}) $
            \IF{ $({\mathcal{G} \notin \boldsymbol{B}_{\text{*}}} \And {l_i \in \mathbf{D}^{'}_{\text{gsar}}})$}
                \FOR{each $i$ in $N_\text{r}$}
                \STATE Attach $\overrightharp{I}^{\text{gsar}}_i$ to model $\mathcal{A}_a$ (Avatar pose recovery for the GSAR)
                \ENDFOR
            \ELSE
                \FOR{each $i$ in $N_{a}$}
                % \STATE update $l_i$ according to Eq. (\ref{eq_l12}) and Eq. (\ref{eq_l11})
                \STATE update $\overrightharp{l}_i$ according to Eq. (\ref{eq_l12}) and Eq. (\ref{eq_l11})

                \STATE Attach $\overrightharp{I}^{\text{egsar}}_i$ to model $\mathcal{A}_a$ (Avatar pose recovery for the E-GSAR)
                \ENDFOR
            \ENDIF
            % \STATE Generate avatar $\hat{A}$ with appearance mesh $\mathcal{A}_a$ and model initial position $l_a$ according to Eq. (\ref{recoverpos}).
            \STATE Generate avatar $\hat{\mathcal{A}_a}$ with appearance mesh $\mathcal{M}_a$ and model initial position $l_a$ according to Eq. (\ref{recoverpos}).

		\ENSURE  Avatar $\hat{\mathcal{A}_a}$ with reconstructed pose
	\end{algorithmic}  
\end{algorithm}
\subsection{Evaluation Metric}
The semantic level of our proposed GSAR aims to enhance the communication effectiveness to achieve accurate avatar moving of the AR application, specifically, the skeleton information accuracy between the transmitter and the receiver. The optimization seeks to minimize the Euclidean distance of the semantic information transmitted at the transmitter and received at receiver.
Thus, the MPJPE is used to estimate and evaluate the avatar pose error in geometry aspect between the transmitter and receiver, including the x-axis, y-axis, and z-axis values, which can be expressed as
% \begin{equation}
% \begin{aligned}
% & \text { MPJPE }=\frac{1}{N_{\mathrm{a}}} \sum_{i=1}^{N_{\mathrm{a}}} \sqrt{\left(\overrightharp{l}_i-\overrightharp{l}^{'}_{i}\right)^2} \\
% & =\frac{1}{N_{\mathrm{a}}} \sum_{i=1}^{N_{\mathrm{a}}} \sqrt{\left(l_{\mathrm{x}}^i-{l}_{\mathrm{x}}^{i'}\right)^2+\left(l_{\mathrm{y}}^i-{l}_{\mathrm{y}}^{i'}\right)^2+\left(l_{\mathrm{z}}^i-{l}_{\mathrm{z}}^{i'}\right)^2},
% \end{aligned}
% \end{equation}
\begin{equation}
\begin{aligned}
& \text { MPJPE }=\frac{1}{N_{\mathrm{a}}} \sum_{i=1}^{N_{\mathrm{a}}} \sqrt{{|\overrightharp{l}_i-\overrightharp{l}^{'}_{i}|}^2},
\end{aligned}
\end{equation}
% \begin{equation}
% \small
% \text{MPJPE} = \frac{1}{N_\text{a}} \sum_{i=1}^{N_\text{a}} \sqrt{(l^{i}_\text{x} - \hat{l}^{i}_\text{x})^2 + (l^{i}_\text{y} - \hat{l}^{i}_\text{y})^2 + (l^{i}_\text{z} - \hat{l}^{i}_\text{z})^2},
% \end{equation}
where the $\overrightharp{l}_i$ and $\overrightharp{l}^{'}_i$ represent the three dimensional position value of skeleton at the transmitter and the receiver respectively.

% \textbf{MPJP}: We use the standard MPJPE loss to estimate the 3D poses of the human and evaluate the avatar pose error between the source and recovered video.
\section {Effectiveness Level Design}
In this section, we will demonstrate the design principles of GSAR optimization at the effectiveness level based on the above defined semantic information. In the following, we present goal-oriented semantic wireless communications and its evaluation metric. 

% \subsection{Channel State Exploration}

\begin{figure}
\centering
\includegraphics[width=7.5cm]{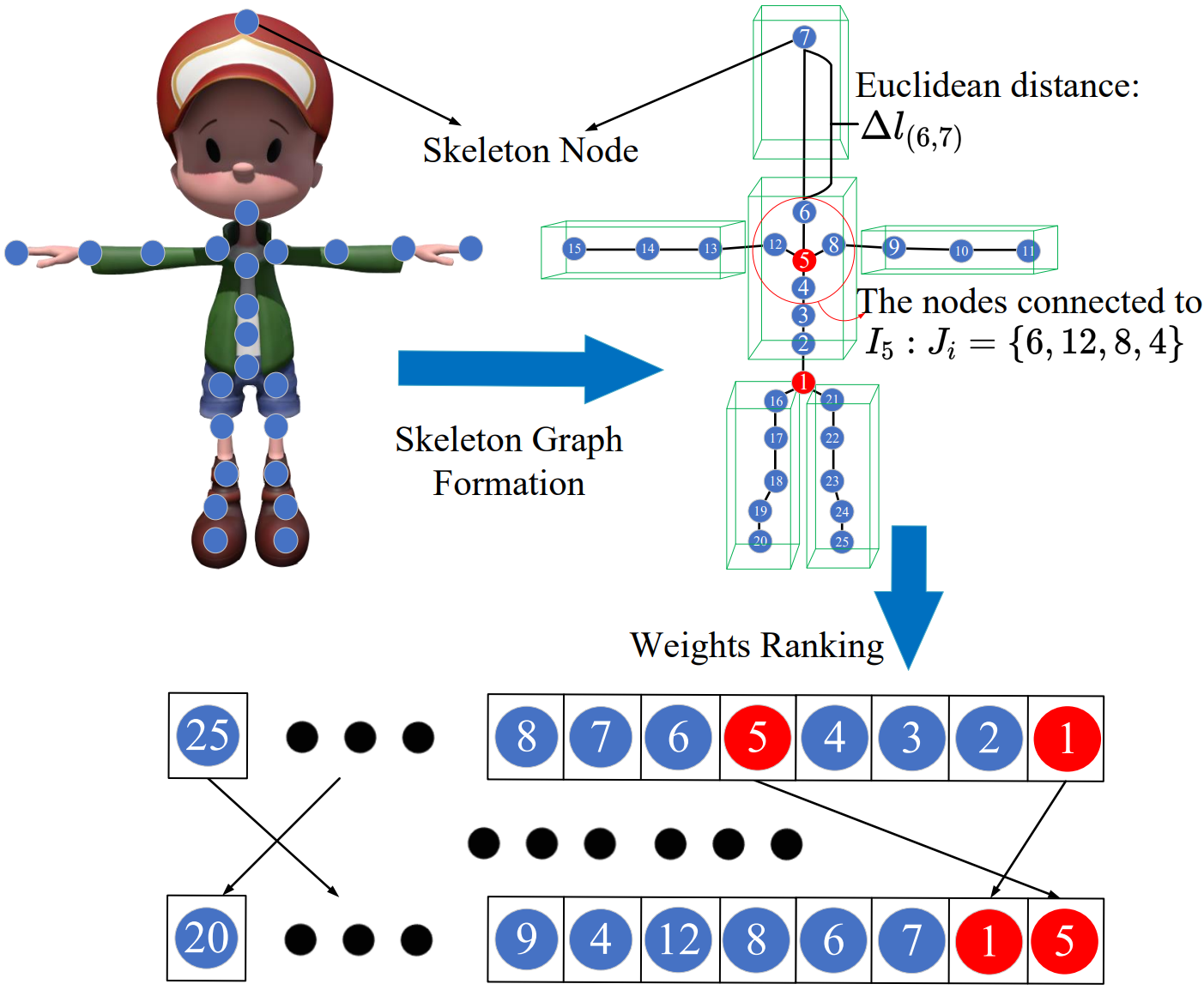}
% [height=4.5cm]表示高度
%[width=9.5cm]表示宽度
%{111.eps}表示eps格式的图片，名为111
\caption{Skeleton graph formation and ranking}
%图片的名称
\label{rankingFigprocesss}
%图片的标签，用于文章中的引用，注意到标签的数字与实际文章显示的数字可能不同
\end{figure}

\subsection{Goal-oriented Semantic Wireless Communications}
To further enhance the effectiveness of avatar communication in AR applications, we 
% have thoroughly investigated current research on avatar models, graph knowledge, and 
propose an avatar-based semantic ranking algorithm to calculate an importance weight value among all the extracted semantic information, which plays a more advantageous role in avatar representation. More specifically, we calculate the importance of the skeleton nodes in the skeleton graph $\mathcal{G}$ using a ranking method based on the 
% ranking method based on the 
PageRank algorithm proposed by Google \cite{joshi2018google}, the detailed process of AbSR algorithm is proposed in $\textbf{Algorithm \ref{avatar_basedRanking}}$, and the weight is calculated as
\begin{equation}
\label{WIcalculate}
\mathrm{\omega}_{I_i}=\frac{N_J}{(1-\alpha)}+\sum_{j=0}^{N_J}\left({|\Delta \overrightharp{l}_{(i,j)}|} \times \omega_{J_j}\right).
\end{equation}
% where $\omega_{I_i}$ represents the weight of the semantic information $\overrightharp{I}_i$ in the $i^{th}$ skeleton node of skeleton graph, and $|\Delta \overrightharp{l}_{(i,j)}|$ denotes the Euclidean distance between the $i^{th}$ and $j^{th}$ skeleton. $\omega_{J_j}$ is the weight value of skeleton $J_j$ connected to skeleton $I_i$, and $\alpha$ is a discount factor ranging from $0$ to $1$. As suggested in \cite{srivastava2017discussion}, we set the discount factor to 0.7 in this paper. $N_\text{j}$ represents the total number of nodes $J_j$ connected to $I_i$ in the graph. 
where $\omega_{I_i}$ represents the weight of the semantic information $\overrightharp{I}_i$ in the $i$-th skeleton node of skeleton graph, and $|\Delta \overrightharp{l}_{(i,j)}|$ denotes the Euclidean distance between the $i$-th and $j$-th skeleton. $J_j$ denotes the node index which are connected to the $i$-th node, $\omega_{J_j}$ is the weight value of the ${J_j}$-th skeleton, $N_\text{j}$ represents the total number of nodes $J_j$ in the skeleton graph, and $\alpha$ is a discount factor ranging from $0$ to $1$. As suggested in \cite{srivastava2017discussion}, we set the discount factor to 0.7 in this paper.
A detailed diagram is shown in Fig. \ref{rankingFigprocesss}, which illustrates that skeletons with more connections and longer distances from other connected skeletons are more critical. The underlying rationale is that a node with more connections will have a greater impact on connected skeleton nodes if it have bit error in wireless communication. Furthermore, nodes that are more isolated, indicated by their greater distance from other skeletons, are likely to have a more substantial impact on the avatar representation due to their distinctive appearance contributions, highlighting the importance of these skeletons.

After calculating the critical node weight of skeleton graph, a descending sort algorithm is applied to arrange the skeleton nodes in descending order of rank. \textcolor{black}{Leveraging our proposed AbSR algorithm, we consider the effectiveness level optimization during the wireless communication, 
% transform the wireless communication process to a task-oriented level,
focusing on avatar semantic preservation. This shift advancing the semantic level design in Section III, thus ensuring that crucial avatar semantic information is prioritized in our goal-based wireless communication approach.} As shown in Eq. (\ref{equationmap}), this approach maps higher weight semantic information to transmit in FDM subchannels with better CSI.
% ultimately enhancing avatar reconstruction capabilities and improving AR application QoE for clients. 
% In this way, we define a new ablation experiment named channel-based TSAR framework (EC-TSAR), with details describing below. 
This is the so called euler angle and channel-based GSAR framework (EC-GSAR), with details below. 

\textbf{EC-GSAR:} 
\textcolor{black}{
Based on the E-GSAR, the CSI information is considered to implement the AbSR algorithm and channel mapping in $\textbf{Algorithm \ref{avatar_basedRanking}}$ to improve communication effectiveness in AR applications. More specifically, the channel mapping process aims to assign more important semantic information a higher priority with the better subchannel for wireless transmission. The semantic information is defined as the vector position and euler angle rotation of all skeletons in the moving avatar, as shown in Eq. (\ref{semanticsinforationetasr}), while the base knowledge encompasses the avatar skeleton graph, shared background model, moving avatar model, and their appearance meshes, as shown in Eq. (\ref{baseknowledgedaiG}).
% Based on the E-TSAR, the CSI information is considered to implement the AbSR and channel mapping in $\textbf{Algorithm \ref{avatar_basedRanking}}$ to improve communication effectiveness in AR applications. More specifically, the channel mapping process aims to assign more important semantic information a higher priority with the better subchannel for later transmission, which is calculated based on the Asbr algorithm. The semantic information is defined as the vector position and Euler angle rotation of all skeletons in the moving avatar, as shown in Eq. (\ref{semanticsinforationetasr}), while the base knowledge encompasses the avatar skeleton graph, shared background model, moving avatar model, and their appearance meshes, as shown in Eq. (\ref{baseknowledgedaiG}).
}
% Based on the E-TSAR, the CSI information is considered to implement the AbSR and channel mapping in $\textbf{Algorithm \ref{avatar_basedRanking}}$ to improve communication effectiveness in AR application. The semantic information is defined as the vector position and euler angle rotation of all skeletons in the moving avatar as shown in Eq. (\ref{semanticsinforationetasr}), while the base knowledge encompasses the avatar skeleton graph, shared background model, moving avatar model, and their appearance meshes, as shown in Eq. (\ref{baseknowledgedaiG}).

\begin{algorithm}[t]
	%\textsl{}\setstretch{1.8}
	\renewcommand{\algorithmicrequire}{\textbf{Input:}}
	\renewcommand{\algorithmicensure}{\textbf{Output:}}
	\caption{Avatar-based Semantic Ranking Algorithm}
    \label{avatar_basedRanking}
	\begin{algorithmic}[1]
		\STATE Initialization: Base Knowledge $\boldsymbol{B}_{\text{*}}$, Semantic \\ information $\mathbf{D}_{\text{gsar}}$
		\STATE  Get $\mathcal{G}, \mathcal{A}_a$ from $\boldsymbol{B}_{\text{*}}$, 
            \STATE  Get $\Delta \overrightharp{l}_{(i,i-1)}$ from $\mathcal{A}_a$
            \STATE  Count skeleton number $N_\text{a} =\mathbf{C}_\text{s}(\mathcal{G})$
            \REPEAT
            % \FOR{each $i$ in $N_{a}$}
            \STATE $k = k + 1$
            \FOR{each $i$ in $N_{\text{a}}$}
                \STATE Update $\omega^k_{I_i} $ with $ \Delta \overrightharp{l}_{(i,i-1)}$ based on Eq. (\ref{WIcalculate})
                \STATE $\delta=||\omega^k_{I_i}-\omega^{k-1}_{I_i}||$
            \ENDFOR
		\UNTIL $  \delta < \varepsilon $  
		\STATE   Update $\{\overrightharp{I}^{\text{gsar}}_i,h_j\}$ according to Eq. (\ref{equationmap})  
		\ENSURE  Channel Mapping $\{ \overrightharp{I}^{\text{gsar}}_i,h_j\}$
	\end{algorithmic}  
\end{algorithm}

\begin{figure*}[t]
\centering
\subfigure[Avatar movement range of adjacent frame.]
{
\includegraphics[width=7.7cm,height=5.6cm]{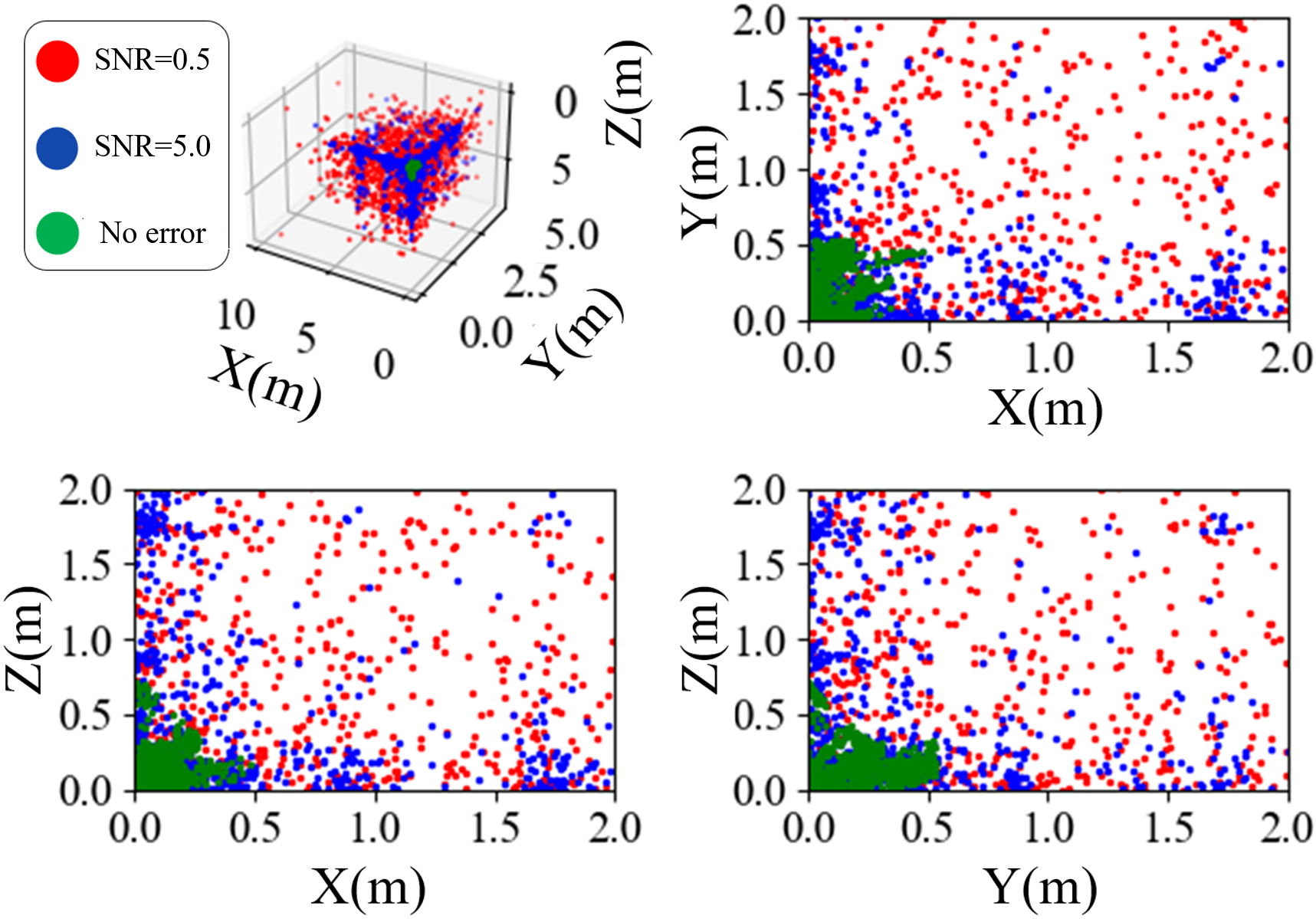}
}
\quad
\subfigure[Semantic information extraction accuracy.]{
\includegraphics[width=7.7cm]{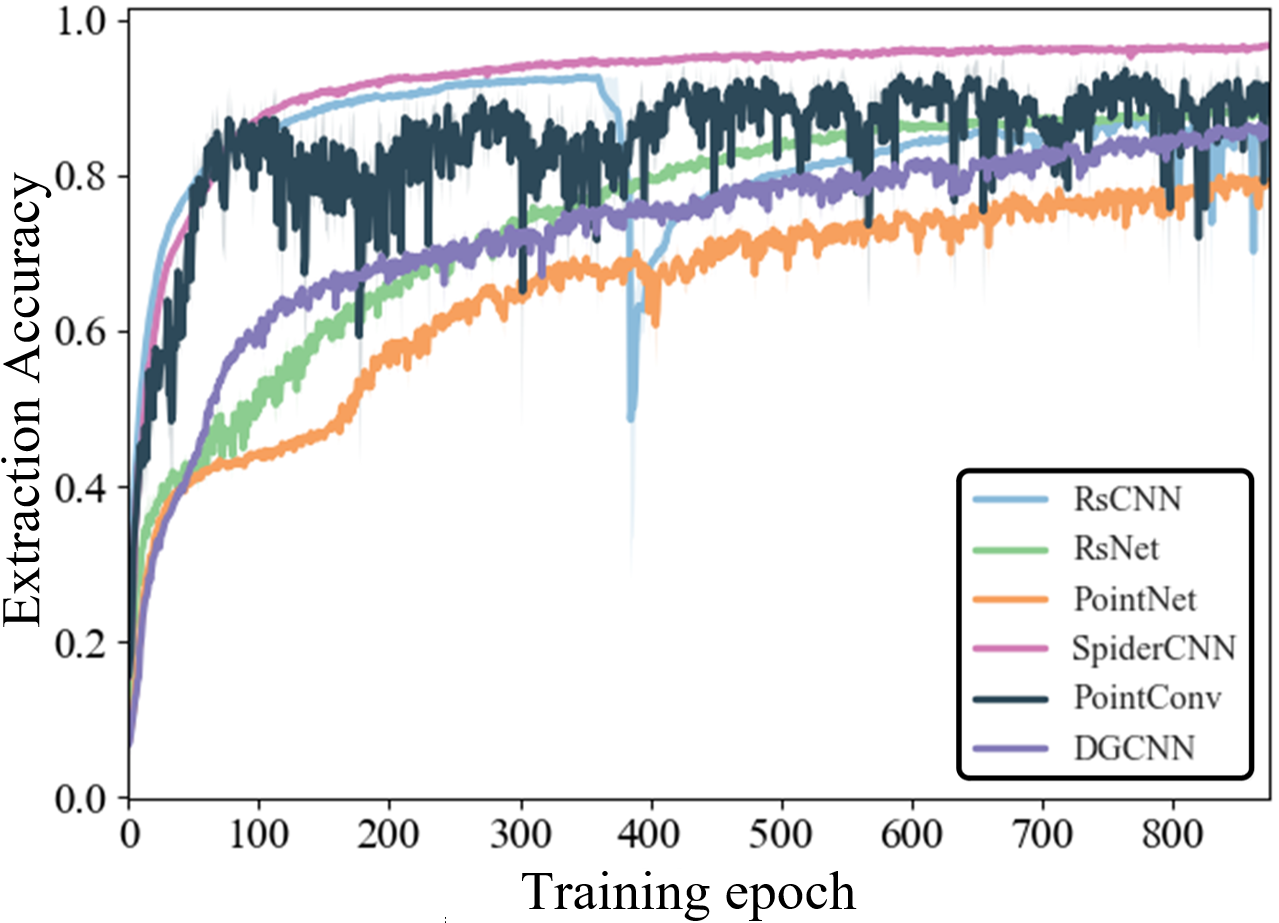}
}
\caption{Avatar movement distribution and semantic information extraction accuracy}
\label{ex11}
\end{figure*}

\subsection{Evaluation Metric}
Building upon semantic level optimization, the overall goal of the goal in AR application is to recover the avatar for better clients viewing experience. To achieve this, we use point cloud to evaluate the entire virtual scenery, which includes Point-to-Point (P2Point), peak signal-to-noise ratio for the luminance component ($\text{PSNR}_\text{y}$), and transmission latency.

\textbf{P2Point} \cite{mekuria2016evaluation}: To evaluate the viewing experience of clients in AR applications, the P2Point metric is employed to assess the AR scenery from a $360^{\circ}$ viewing angles, comparing the geometry difference between the point cloud data at transmitter $\mathbf{P}_{\text{t}}$ and the point cloud data at receiver $\mathbf{P}_{\text{r}}$.
% Given the cloud points at receiver 
% $P_{\text{r}}=\{v_1, \ldots, v_n\}$, 
% P2Point calculates the geometric difference from the cloud points at transmitter $\mathbf{P}_{\text{t}}$. 
The P2Point error calculation can be expressed as
\begin{equation}
\text{P2Point}= \max \left({d}_{\text{rms}}^{\left(\mathbf{P}_{\text{t}}, \mathbf{P}_{\text{r}}\right)}, {d}_\text{rms}^{\left(\mathbf{P}_{\text{r}}, \mathbf{P}_{\text{t}}\right)}\right),
\end{equation}
where the function ${d}_{\text{rms}}$ is the root mean square error between two point cloud.

$\textbf{PSNR}_\textbf{y}$ \cite{meynet2020pcqm}: The color difference plays a crucial role in avatar displaying AR applications, as it can significantly impact the user viewing experience if there are discrepancies in the colors transmitted. The $\text{PSNR}_\text{y}$ is used to evaluate the luminance component of the AR scenery difference between the receiver and transmitter
% from a $360^{\circ}$ perspective
. The $\text{PSNR}_\text{y}$ is then calculated as
% \textcolor{black}{
% \begin{equation}
% \mathbf{d}_\text{y}(\mathbf{P}_{\text{t}},\mathbf{P}_{\text{r}})=\sqrt{\frac{1}{m} \sum_{\overrightharp{v} \in \mathbf{P}_{\text{t}}}\left[\mathbf{y}(\overrightharp{v})-\mathbf{y}(\mathbf{d}_\text{near}(\overrightharp{v},\mathbf{P}_{\text{r}}))\right]^2},
% \end{equation}
\begin{equation}
\text{PSNR}_\text{y}=10 \log_{10}\left(\frac{255^2}{{\frac{1}{N_\text{t}} \sum_{\overrightharp{v}_i \in \mathbf{P}_{\text{t}}}\left[{y}_{\overrightharp{v}_i}-{y}_{\overrightharp{v}^{\mathbf{P}_{\text{r}}}_\text{near}}\right]^2}}\right),
\end{equation}
% }
where $\overrightharp{v}_\text{near}^{\mathbf{P}_{\text{r}}}$ represents the nearest point to $\overrightharp{v}_i$ from point cloud $\mathbf{P}_{\text{r}}$, $N_\text{t}$ represents the total number of point cloud in the $\mathbf{P}_\text{t}$, and ${y}_{\overrightharp{v}_i}$ represents the luminance elements of point $\overrightharp{v}_i$.
% \begin{equation}
% \text{PSNR}_\text{y}=10 \log_{10}\left(\frac{255^2}{\mathbf{d}_\text{y}(\mathbf{P}_{\text{t}}, \mathbf{P}_{\text{r}})^2}\right).
% \end{equation}

\textbf{Transmission Latency}: Transmission Latency is a critical metric in AR applications and plays a crucial role in evaluating client QoE. The transmission latency of the AR application can be divided into different components, including semantic information extraction time  $T_\text{s}$, wireless communication time $T_\text{w}$, avatar pose recovery and rendering time $T_\text{r}$. The combination of all these times results in the transmission delay of the AR application, which can be expressed as
\begin{equation}
\label{time_delay}
\text{Transmission Latency}=T_\text{s}+T_\text{w}+T_\text{r},
\end{equation}
by analyzing and optimizing each component of the transmission latency, we can justify and indicate the efficiency of our proposed framework.
% ensuring a smoother and more immersive experience for users while maintaining high-quality communication between the transmitter and receiver.

\section {Simulation Results}
\textcolor{black}{
In this section, we evaluate the performance of our proposed GSAR framework and compare it with the traditional point cloud communication framework as well as the enhanced frameworks such as E-GSAR and EC-GSAR, as described in sections III and IV.
To assess the performance of semantic information extraction, we utilize several different avatar dance types as specified in Table I. We also configure the hyperparameters for the SANet and wireless communication as listed in Table II. This includes the learning rate, batch size, channel fading type, base knowledge information, and so on. The experimental platform for this study employs Python 3.9.0 on the Ubuntu 18.04 system with an RTX 3070, PyTorch 2.1, and the Unity platform.
The SANet initially undergoes a learning phase where it is trained until it converges to an optimal state. Once the training phase is complete, the trained neural network is implemented across GSAR, E-GSAR, and EC-GSAR.
The following sections present the results of our proposed frameworks. Section V-A offers insights into the avatar movement distribution and Section V-B first provides the experiment results on the semantic information extraction accuracy achieved by the SANet. Following that, we present experimental results examining various metrics to evaluate the XR application and avatar transmission. These metrics include the mean per joint position error (MPJPE), the adjacent frame MPJPE, transmission latency, Point-to-Point (P2Point) error, and peak signal-to-noise ratio ($\text{PSNR}_y$).
}

\subsection{Avatar Skeleton Distribution}
\begin{table}[t]
  \centering
  \caption{Experiment Setup}
  \label{tab:Experiment_Setup}
  \begin{tabular}[l]{@{}lc}
  \toprule
       \textbf{\emph{Dance type}}  & \textbf{\emph{Last time}}\\ 
      Upper body dance &  2min 10s \\
       Slight shaking &   50s \\ 
      Full body dance &  2min 5s \\ 
      \midrule
      
      \textbf{\emph{Simulation}}  & \textbf{\emph{Value}}\\
        Data type &  Point cloud \\ 
        FPS &  60 Hz \\ 
        Avatar skeleton number &  25 \\ 
        Stationary model skeleton number&  15 \\ 
        Point cloud number&  2,048 \\ 
       Attribute information 1 &  Point number \\ 
      Attribute information 2&  Position \\ 
      Attribute information 3 &  Rotation (optional) \\ 
     Attribute information 4 &  Color (optional) \\ 
      \textcolor{black}{Channel response} &  \textcolor{black}{Rayleigh fading} \\ 
             \textcolor{black}{Modulation} &  \textcolor{black}{BPSK} \\ 
             \textcolor{black}{Location movement range (x,y,z)} &  \textcolor{black}{([0,0.52], [0,0.76], [0,0.74])}\\ 
      \midrule

      \textbf{\emph{\textcolor{black}{Base knowledge}}}  & \textbf{\emph{\textcolor{black}{Symbols}}}\\
        \textcolor{black}{Avatar skeleton graph} & \textcolor{black}{$\mathcal{G}$} \\ 
        \textcolor{black}{Avatar initial position} &  \textcolor{black}{$l_o$} \\ 
        \textcolor{black}{Avatar model} &  \textcolor{black}{$A_a$} \\ 
        \textcolor{black}{Stationary background model}  &  \textcolor{black}{$A_s$} \\ 
        \textcolor{black}{Stationary initial position} &  \textcolor{black}{$l_s$} \\ 
       \textcolor{black}{Appearance meshes} & \textcolor{black}{$M_a$, $M_s$} \\ 
    \bottomrule
  \end{tabular}
\end{table}
% \begin{table}[t]
%   \centering
%   \caption{Experiment Setup}
%   \label{tab:Experiment_Setup}
%   \begin{tabular}[l]{@{}lc}
%   \toprule
%        \textbf{\emph{Dance type}}  & \textbf{\emph{Last time}}\\ 
%       Upper body dance &  2min 10s \\
%        Slight shaking &   50s \\ 
%       Full body dance &  2min 5s \\ 
%       \midrule
      
%       \textbf{\emph{Simulation}}  & \textbf{\emph{Value}}\\
%         Data type &  Point cloud \\ 
%         FPS &  60 \\ 
%         Avatar skeleton number &  25 \\ 
%         Stationary model skeleton number&  15 \\ 
%         Point cloud number&  2,048 \\ 
%        Attribute information 1 &  Point index \\ 
%       Attribute information 2&  Position \\ 
%       Attribute information 3 &  Rotation (optional) \\ 
%      Attribute information 4 &  Color (optional) \\ 

%     \bottomrule
%   \end{tabular}
% \end{table}

To obtain a comprehensive understanding of avatar movement in the AR environment, several avatar dance types were conducted upon the Unity3D and Mixamo platform. Mixamo is a robust 3D character creation and animation tool offering a wide array of diverse and dynamic 3D character animations suitable for a broad spectrum of movement analysis. Three distinct dance types from Mixamo were selected for our experiments: an upper-body dance, a slight shaking dance, and a full-body dance. These dances cover a wide range of avatar movements, from localized to full-body motions, and each dance has a specific duration, as detailed in Table \ref{tab:Experiment_Setup}. The transmitter used for these experiments operates at 60 frames per second (FPS), ensuring a smooth and continuous displaying of the avatar's movements at the transmitter. 
% The resulting data from these movements, represented as point cloud data, contain attributes such as point index, position, rotation (optional), and color (optional). 
% This data is stored in a JSON label file for subsequent processing. 
The moving avatar, with 25 skeletons, is placed on a stationary background stage model.

\textcolor{black}{Fig. \ref{ex11} (a) plots the data analysis of the experiments, which is carried out based on the skeleton difference between the adjacent frames across the X, Y, and Z axes under different SNR sceneries. Green points correspond to adjacent frame skeleton position differences under optimal wireless channels, which reveals that the shifts in position from one frame to the next were typically minimal. \textcolor{black}{The adjacent difference ranges for the three axes are (0, 0.52), (0, 0.76), and (0, 0.74) meters, respectively, suggesting that the maximum movement of the avatar's skeleton usually remains from 0.52 to 0.76 meters} per frame in the Unity3D platform.
Furthermore, with the SNR increases, the adjacent skeleton difference indicates that the received data might be distorted under highly noisy conditions and the Rayleigh fading channel. This can result in significant positional differences between adjacent frames, potentially surpassing the realistic movement capabilities of the avatar and subsequently causing disjointed in the virtual environment.}

\begin{figure*}[t]
\centering
\subfigure[Adjacent MPJPE of GSAR.]
{
\includegraphics[width=5.3cm]{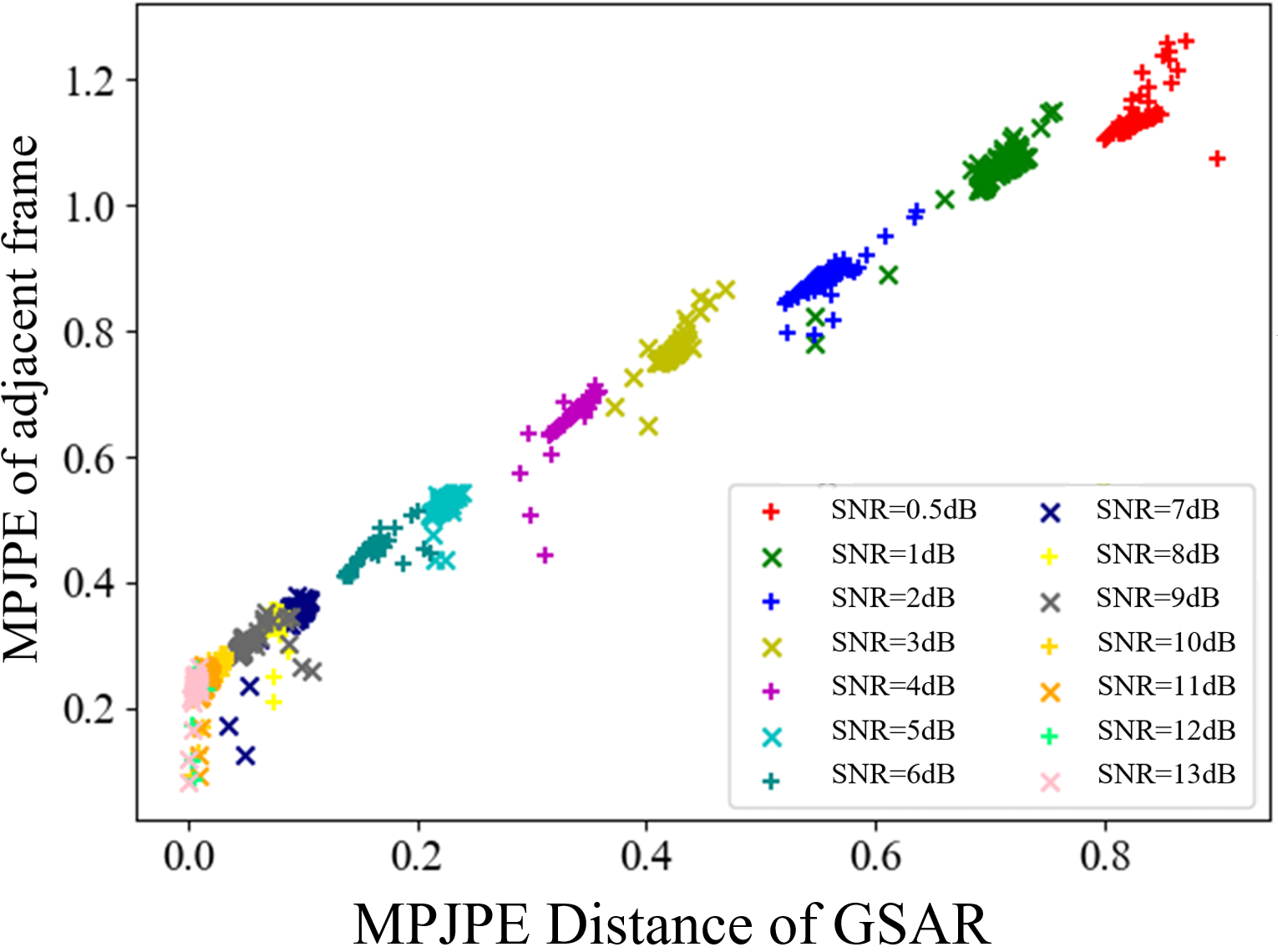}
}
\quad
\subfigure[Adjacent MPJPE of E-GSAR.]{
\includegraphics[width=5.3cm]{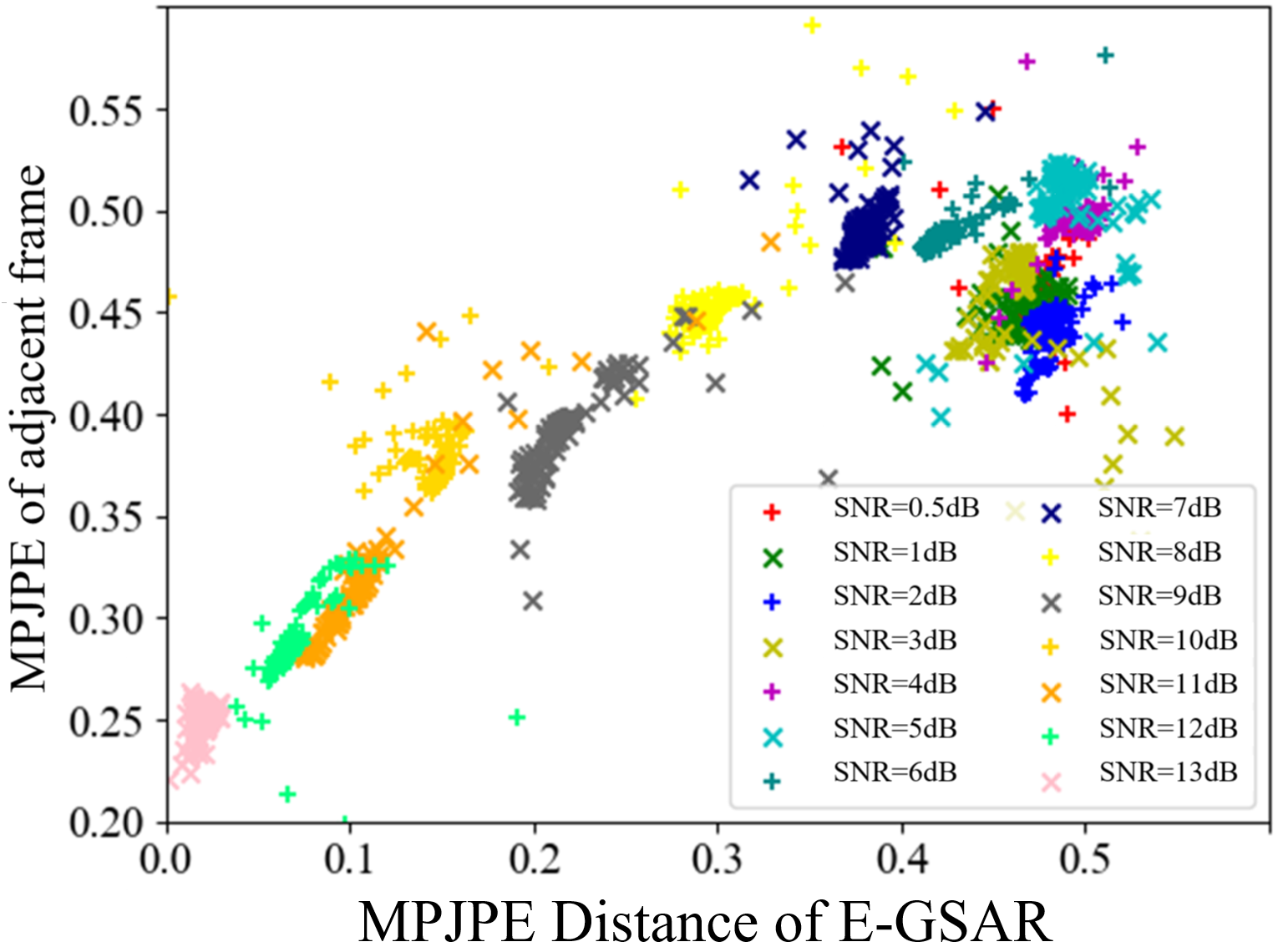}
}
\quad
\subfigure[Adjacent MPJPE of EC-GSAR.]{
\includegraphics[width=5.3cm]{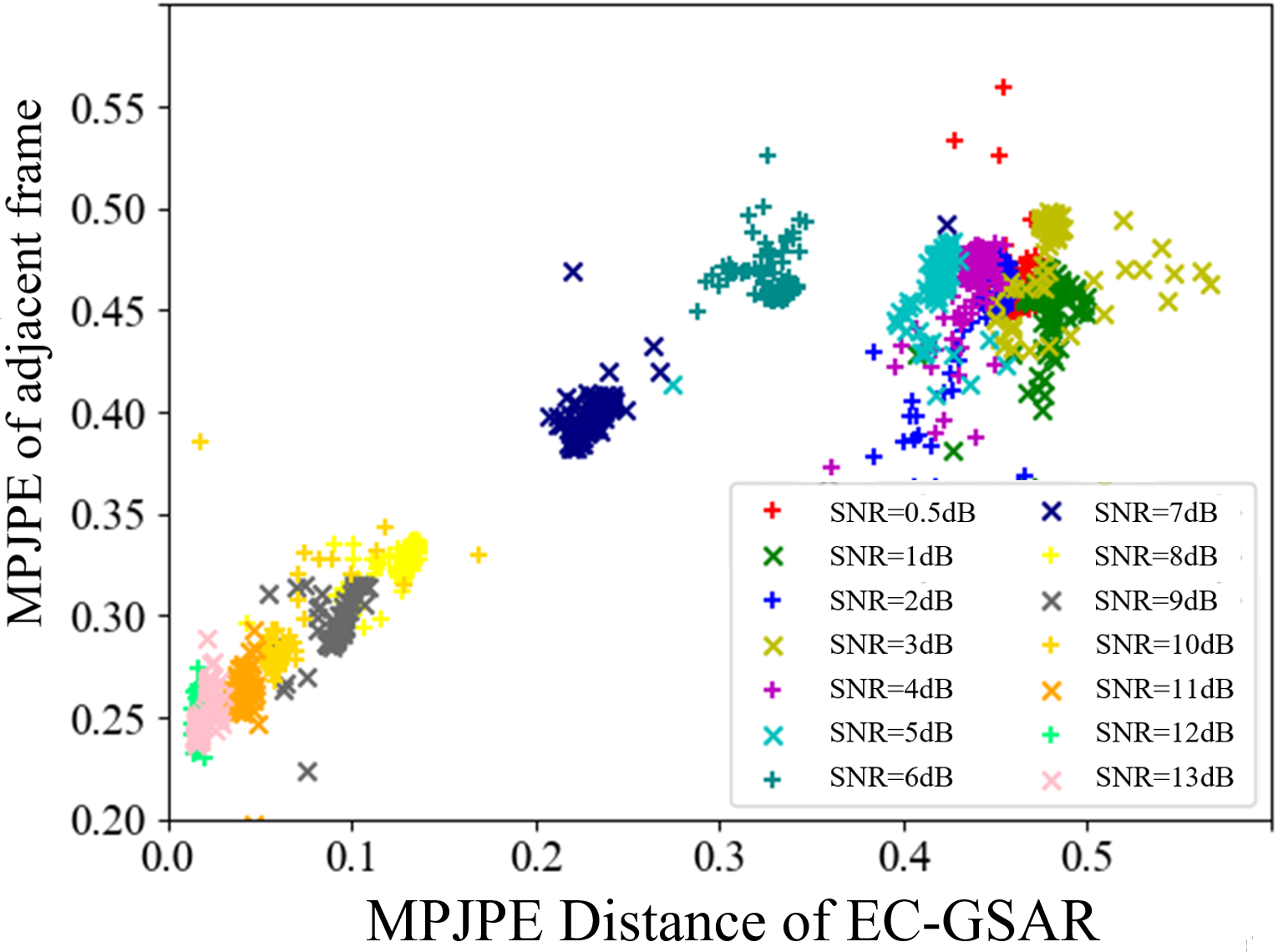}
}
\quad
\caption{Adjacent MPJPE difference among GSAR, E-GSAR, and EC-GSAR}
\label{ex22}
\end{figure*}

\subsection{Performance Evaluation}
\subsubsection{Semantic information Extraction Performance}
% \subsection{semantics Extraction Performance}
% Fig. \ref{Fig_2Training_results} demonstrates that the SpiderCNN-based SANet attains the highest accuracy in skeleton information extraction, exceeding the accuracy of over 96\% among the same training epochs. This outcome makes the SANet achieves significantly more reliable and stable compared to other backbone-based networks.
\textcolor{black}{Fig. \ref{ex11} (b) plots the semantic extraction precision of the SANet, anchored on a variety of backbone networks over equivalent training epochs. Each network exhibits commendable proficiency, corroborating the viability of employing such a deep learning mechanism to extract semantic information from point cloud data. The degree of accuracy serves as a benchmark for the effectiveness of semantic extraction capabilities, the accuracy of which is delineated as follows: SpiderCNN \textgreater PointConv \textgreater RsNet \textgreater RsCNN \textgreater DGCNN. This pecking order underscores the pronounced superiority of the SpiderCNN-based SANet, achieving an impressive accuracy surpassing 96\% within the same epoch duration. As outlined in Table \ref{tab:Experiment_Setup}, the SpiderCNN boasts a unique structural design that performs better in point cloud structure feature extraction. This advantage may become particularly obvious in handling complex, high-dimensional data such as avatars and 3D model structures.
This could also illuminate the other backbone networks' less efficient processing and learning capacities. It is likely that other backbones struggle with adequately extracting and learning from the structure of point cloud structure, which could consequently impact semantic information extraction accuracy. These findings highlight the importance of not just the SANet, but also the backbone choice while performing semantic information extraction over point cloud data. 
% Future research may focus on optimizing these networks further and exploring the impact of other backbone architectures on the performance of SANet.
}

% \begin{figure*}[t]
% \centering
% \subfigure[Adjacent MPJPE of TSAR.]
% {
% \includegraphics[width=4.5cm]{AdjMPJP_TASO.png}
% }
% \quad
% \subfigure[Adjacent MPJPE of E-TSAR.]{
% \includegraphics[width=4.4cm]{AdjMPJP_TASO_Eular.png}
% }
% \quad
% \subfigure[Adjacent MPJPE of EC-TSAR.]{
% \includegraphics[width=4.5cm]{AdjMPJP_TASO_CSI.png}
% }
% \quad
% \caption{Adjacent MPJPE difference among TSAR, E-TSAR, and EC-TSAR}
% \label{ex22}
% \end{figure*}

% \subsection{Experiment Evaluation}
\subsubsection{Avatar Transmission Performance}
% \textcolor{black}{To assess the quality of avatar transmission in wireless AR applications, we evaluated both the adjacent MPJPE and its relationship with the MPJPE between the receiver and the transmitter. Generally, the adjacent MPJPE at the receiver, which measures the error of the avatar skeleton between consecutive frames, serves as an indicator of video fluency. On the other hand, the MPJPE between the receiver and the transmitter reflects the accuracy of the avatar's movements.}
% In addition to the MPJPE of the avatar skeleton positions between the transmitter and receiver, the client's QoE also heavily depends on the fluency of the received video. A clear and fluid video transmission is critical to maintaining a high QoE, which ultimately affects the user experience.
% To better understand the influence of the proposed TSAR optimization on both semantics and task levels, we've examined the relationship between two key factors: the MPJPE of the avatar skeleton between adjacent frames at the receiver (indicative of video fluency), and the MPJPE between the receiver and transmitter (representative of the accuracy of the avatar movement). 
\textcolor{black}{Fig. \ref{ex22} (a) plots the MPJPE of adjacent frames, alongside the MPJPE error between the receiver and transmitter, under different wireless channel conditions for the proposed GSAR. With the diminishing SNR, a visible degradation in AR displaying fluency with uncontinued avatar movement of adjacent frames, marked by an increase in both the adjacent MPJPE and the MPJPE. This result reemphasize the insights drawn from Fig. \ref{ex11} (a), signifying that a lower SNR channel generates noise and blur in the received packets, thereby increasing the MPJPE.
Furthermore, with the SNR decrease below 5 dB, the MPJPE of adjacent frames amplifies with the decreasing SNR and transcends the general avatar movement range under optimal wireless channels explicated in section V-A. This demonstrates that concerning the adjacent MPJPE, with the SNR decrease, it alludes to precipitous movements of the avatar's constituent parts, potentially inducing stutters when substantial positional discrepancies arise between successive frames. Simultaneously, if the MPJPE escalates excessively, it could engender distortions in the avatar, with skeletal elements manifesting in aberrant positions, such as a foot emerging at the head. Both the uninfluenced and distortion of the avatar in the AR application could damage the viewing experience on the client side \cite{fiedler2010generic}.}

Fig. \ref{ex22} (b) plots the MPJPE of adjacent frames, alongside the MPJPE error between the receiver and transmitter, under different wireless channel conditions for the proposed E-GSAR. 
In contrast to the outcomes of our proposed GSAR shown in Fig. \ref{ex22} (a), E-GSAR profoundly decreased the MPJPE between the transmitter and the receiver with the SNR increase and achieved a 40\% decrease in MPJPE within the 0.5 dB SNR scenery. Such observations denote a smoother and more fluent avatar movement of the E-GSAR compared to the GSAR, given the E-GSAR a reduced likelihood of confronting disconcerting avatar distortions compared to GSAR. 
% This suggests that although the transmitted data may have a considerable difference compared to the original data, the avatar in the AR system at the receiver side may only generate an odd position, but will not distort the avatar figure like the base TSAR framework. 
% Since the MPJPE under all the SNR scenarios is roughly equal to the maximum avatar movement ability, users at the client side may experience better QoE since they don't need to watch scared distortion avatar. 
Additionally, unlike the basic GSAR results, where the MPJPE continues to increases as the SNR decreases, the E-GSAR MPJPE does not increase after the SNR drops below 5 dB. This indicates that using the avatar model as base knowledge in semantic communications helps the avatar maintain its undistorted appearance in the poor wireless channel scenarios. This improvement in avatar representation can lead to an enhanced user experience and a higher QoE for clients, 
thereby underscoring the effectiveness of employing the avatar model as a shared base knowledge in the domain of wireless AR implementations.
% indicating using the avatar model as shared base knowledge is good in wireless AR displaying application.
% making E-TSAR a promising framework for handling challenging communication conditions for avatar-centric AR application.

\begin{figure}[t]
\centering
\includegraphics[width=7.7cm]{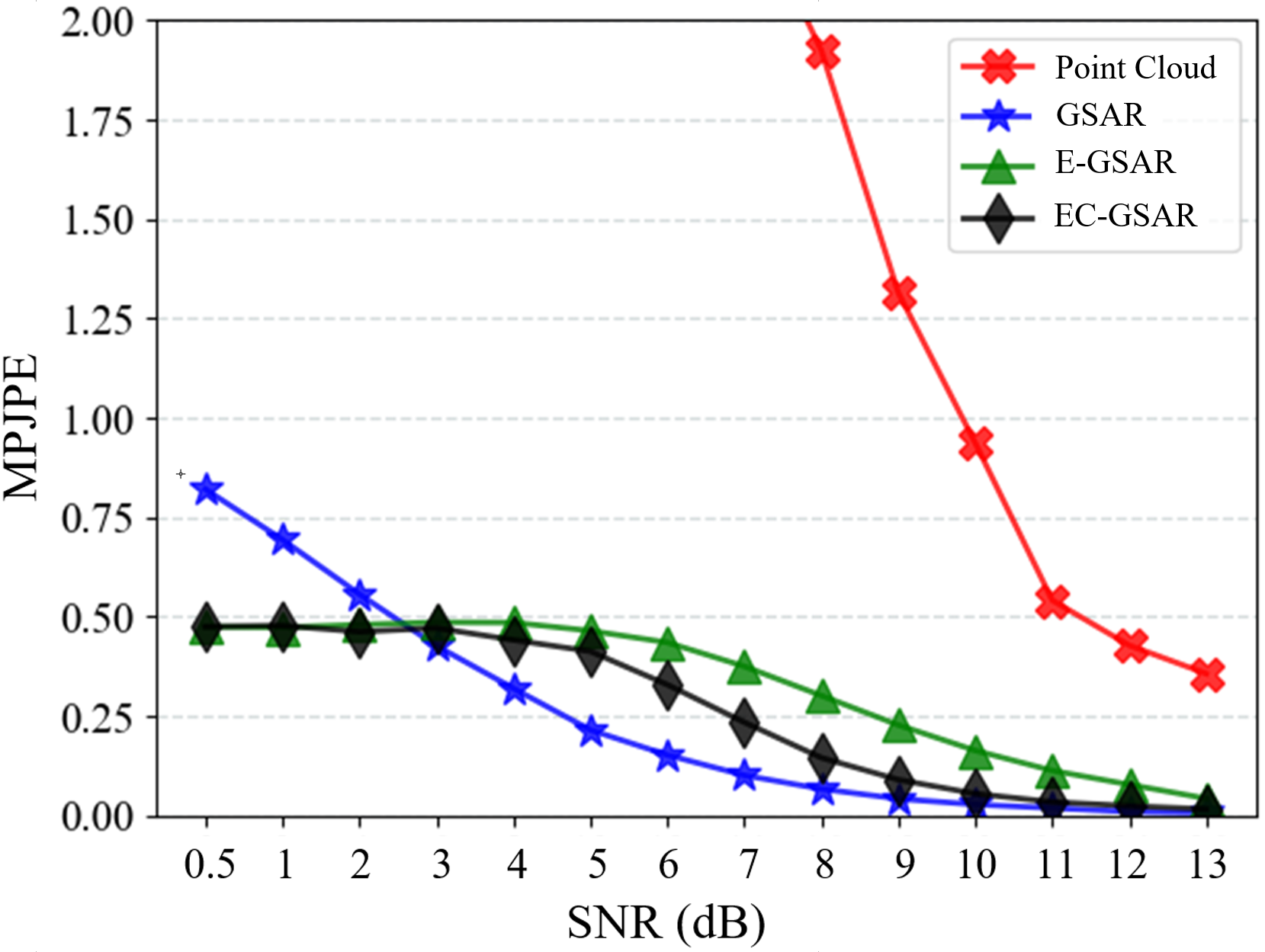}
\caption{Mean Per Joint Position Error.}
\label{ex33}
\end{figure}
% Figure \ref{ex22}(b) plots the MPJPE of adjacent frames, alongside the MPJPE error between the receiver and transmitter, under different wireless channel conditions for the proposed EC-TSAR.  The results are generally similar to those of E-TSAR. However, there is a slight decrease in MPJPE at the worst SNR scenario. Most importantly, EC-TSAR achieves a significant improvement when the SNR is higher than 10 dB. This illustrates that with the assistance of the AbSR algorithm and adaptive channel mapping, the more useful skeleton information \textcolor{black}{is effectively transmitted through wireless communication}, ultimately aiding in avatar recovery on the client side.
% Furthermore, a lower MPJPE of adjacent frames with the EC-TSAR indicates a more fluid video compared to E-TSAR. Similar to the E-TSAR, the MPJPE does not continue to increase as the SNR decreases, unlike the base TSAR. This highlights the outstanding performance of base knowledge selection and the AbSR algorithm in avatar transmission for AR applications. Thus, the EC-TSAR framework demonstrates the effectiveness of the AbSR algorithm and adaptive channel mapping in improving the accuracy of transmitted data, especially in higher SNR scenarios. 
\begin{figure*}[t]
\centering
\subfigure[Point to point.]{
\includegraphics[width=7.7cm]{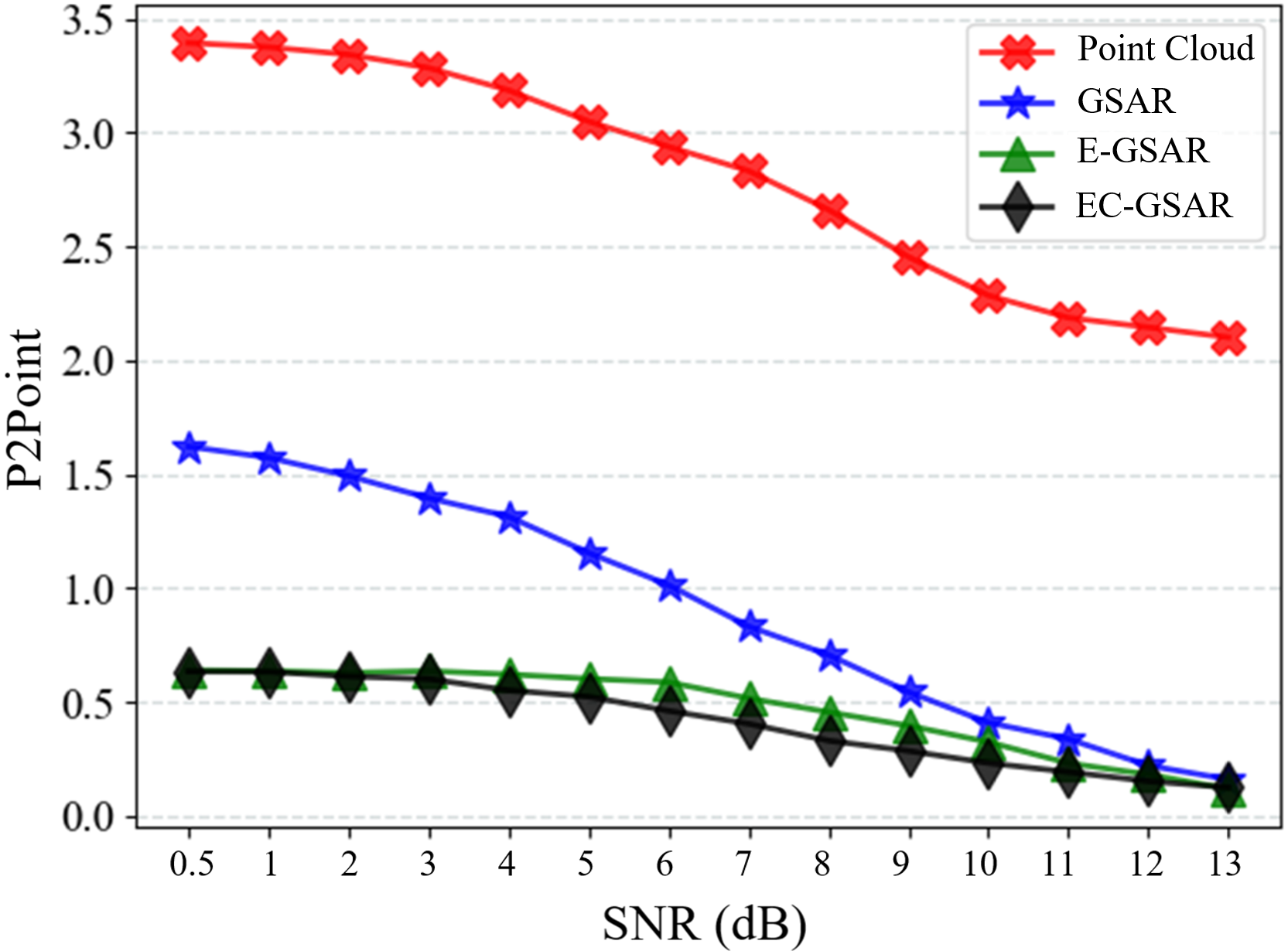}
}
\quad
\subfigure[Peak signal-to-noise ratio in the luminance (Y).]{
\includegraphics[width=7.7cm]{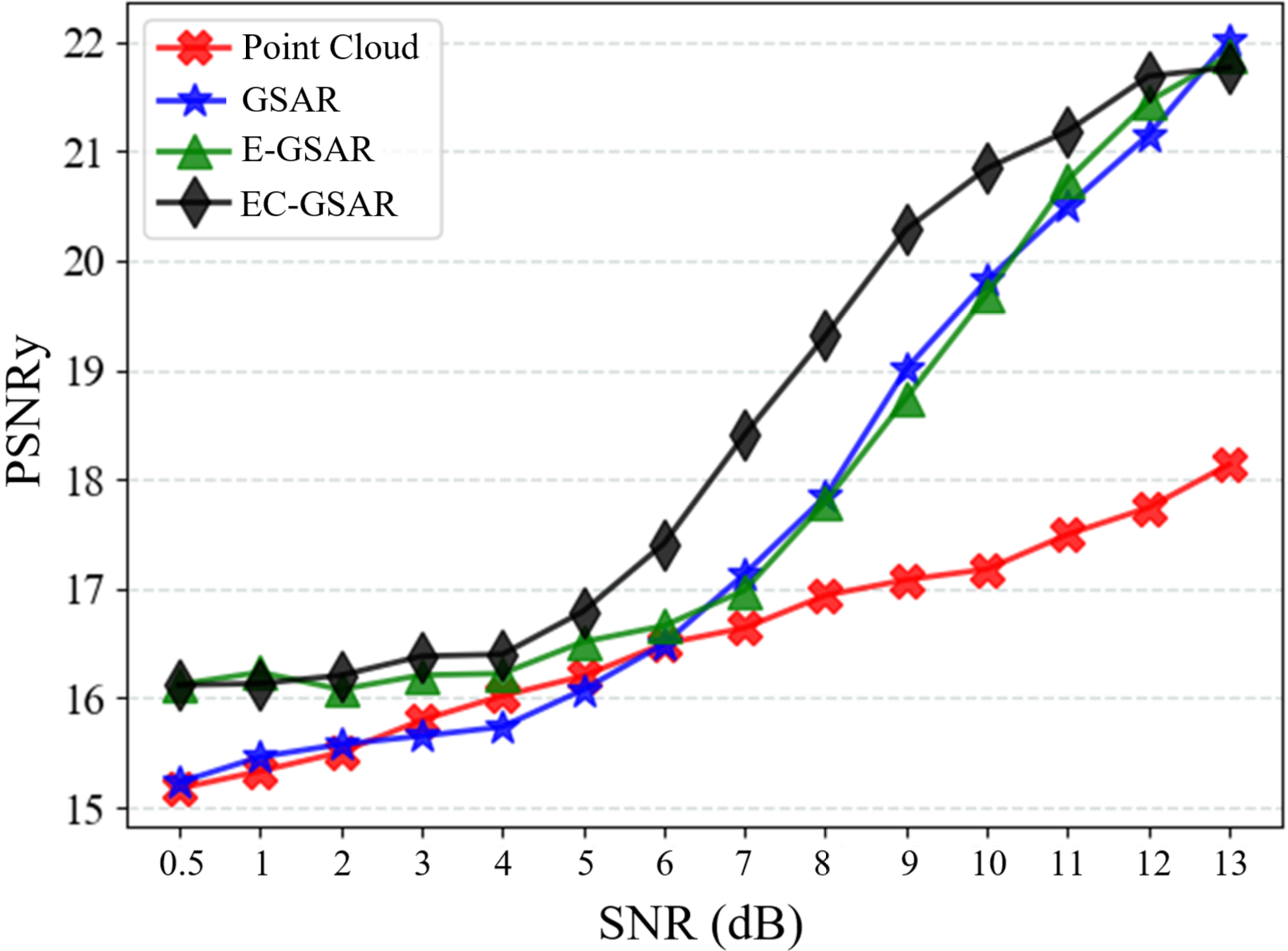}
}
\quad
\caption{Point to point and peak signal-to-noise ratio in the luminance (Y).}
\label{ex55}
\end{figure*}
\begin{figure}[t]
\centering
\includegraphics[width=7.7cm]{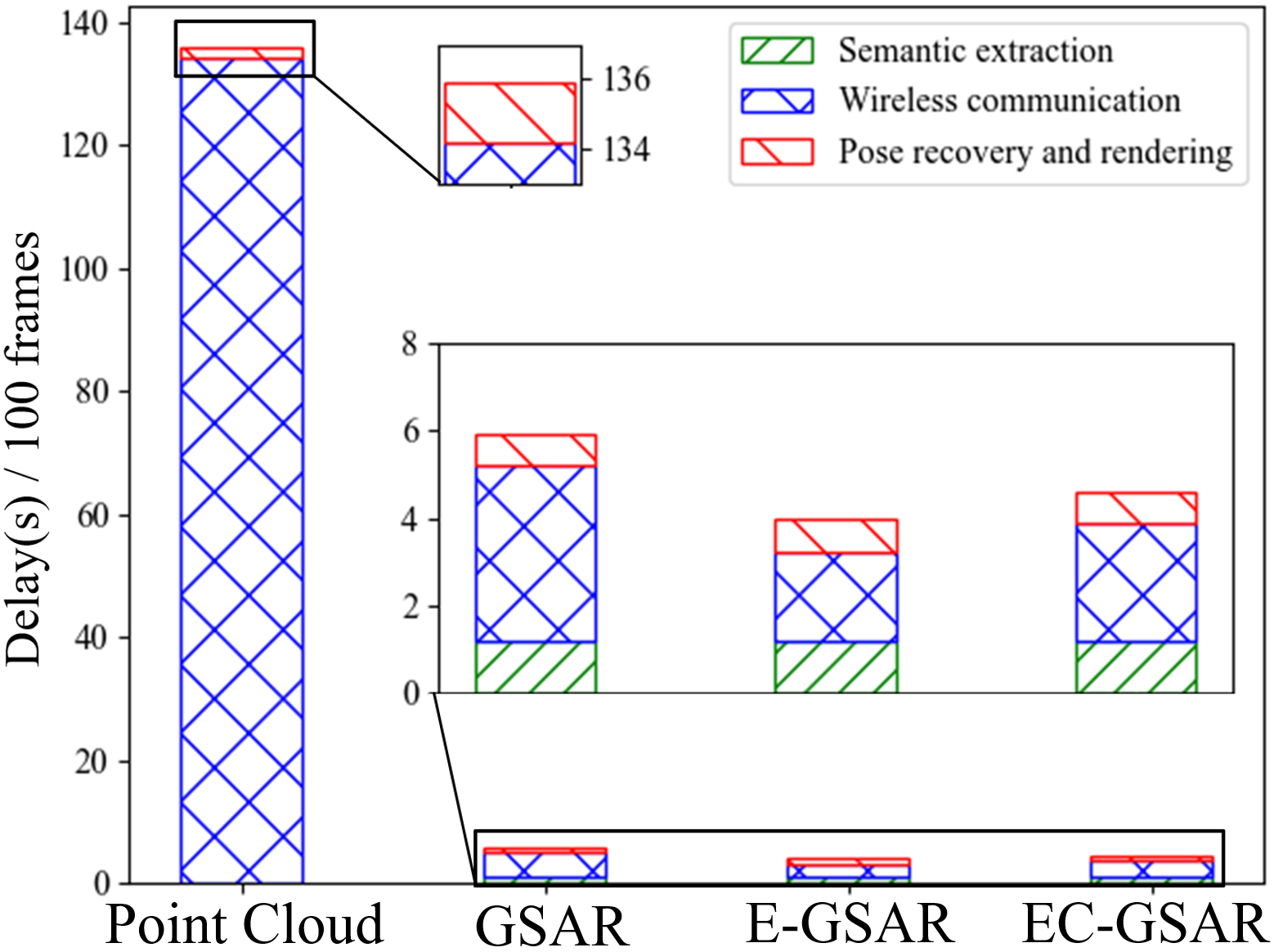}
\caption{Transmission Latency.}
\label{ex44}
\end{figure}

Fig. \ref{ex22} (c) plots the MPJPE of adjacent frames, alongside the MPJPE error between the receiver and transmitter, under different wireless channel conditions for the proposed EC-GSAR. With a result generally similar to E-GSAR's shown in \ref{ex22} (b), EC-GSAR achieves a significant decrease when the SNR increase above 5 dB, generating a more fluent video with lower adjacent frames MPJPE. This illustrates that with the assistance of the AbSR algorithm and adaptive channel mapping, more important semantic information is effectively transmitted through wireless communication, ultimately aiding in avatar recovery on the client side. This highlights the effectiveness of the AbSR algorithm and adaptive channel mapping in improving the efficacy of avatar transmission, especially in higher SNR scenarios. Besides, similar to the E-GSAR, the MPJPE does not continue to increase as the SNR decreases below 5 dB, which reemphasizes the advantages of employing the avatar model as a shared base knowledge.

Fig. \ref{ex33} plots the MPJPE performance results, which reveal the differences in the avatar skeleton's position between the receiver and transmitter. A lower MPJPE indicates a better avatar pose recovery ability in wireless communication, and the overall results of The MPJPE results are ranked as GSAR $\textless$ EC-GSAR  $\textless$ E-GSAR $\textless$ Point Cloud. Specifically, the GSAR framework achieves the lowest MPJPE with the SNR  increase above 3 dB, achieving about an 83\% decrease compared to the point cloud framework at 13 dB scenery. In contrast, the EC-GSAR framework achieves lower MPJPE than the GSAR framework when the SNR continues to decrease below 3 dB. Besides,  the point cloud framework struggles to generate key points within the 3D scenery with the SNR decrease below 8 dB. This observation indicates that in the cloud point communication framework, the avatars are displayed with distorted proportions, such as an arm's length longer than the avatar's entire body, which can cause the SANet to fail in distinguishing the skeleton key points accurately. Meanwhile, in the EC-GSAR, the avatar model used in the shared base knowledge functions not to allow movements exceeding the avatar's capabilities, resulting in a better and undistorted AR avatar displayed on the client side compared with other frameworks with the SNR continue to decrease below 3 dB.

Fig. \ref{ex55} (a) plots the P2Point error, revealing the geometry differences of the AR scene between the transmitter and receiver. A lower P2Point value indicates a better viewing experience of the geometry aspect on the client side, and the overall P2Point value is ranked as EC-GSAR $\textless$ E-GSAR $\textless$ GSAR $\textless$ Point Cloud. With the SNR increases, the P2Point of all the frameworks witnessed an increase, indicating all the frameworks are affected by the worse wireless channel conditions. Besides, The EC-GSAR and E-GSAR frameworks both achieve a flat P2Point value increase with the SNR decrease below 4 dB compared with GSAR and Point Cloud, indicating that the avatar model transmitted in the base knowledge works to prevent the avatar displaying distortion, and
make avatar only generates some odd positions in both frameworks, while the avatar displaying in the point cloud framework and GSAR already shows distortion.

% Fig. \ref{ex55}(a) plots the P2Point results, which reveal the differences in the geometry of the entire AR scene between the transmitter and receiver, not just the avatar. 
% The P2Point value is ranked as Point Cloud $\textgreater$ TSAR $\textgreater$ E-TSAR $\textgreater$ EC-TSAR in all SNR scenarios.
% Specifically,
% as the SNR increases, the point cloud transmission framework continues to increase and exhibits the worst results. In contrast, the EC-TSAR framework yields the best results. The EC-TSAR and E-TSAR frameworks demonstrate similar P2Point performance at both the best and worst wireless channel environments, indicating that the avatar only generates some odd positions in both frameworks, while the point cloud framework in the TSAR method already shows distortion.
% The similar performance of EC-TSAR and E-TSAR at low SNR values likely suggests that the avatar exhibits some unusual positions in both EC-TSAR and E-TSAR but does not display significant distortion as observed in the point cloud framework. 

Fig. \ref{ex55} (b) plots the $\text{PSNR}_\text{y}$ results, which reveal the color differences of the AR displaying scenery between the transmitter and receiver. A higher $\text{PSNR}_\text{y}$ value represents a better viewing experience on the client side, and the $\text{PSNR}_\text{y}$ results are ranked as EC-GSAR $\textgreater$ E-GSAR $\textgreater$ GSAR $\textgreater$ Point Cloud. All the frameworks shown an increase with the SNR increase, indicating the viewing experience is affected by the wireless channel conditions. Besides, all the GSAR, E-GSAR, and EC-GSAR achieve a significant increase when the SNR increase above 7 dB, while the point cloud communication framework has a relatively flat increase. This indicates the avatar model used in the shared base knowledge makes the avatar transmitted as a whole model, which helps to more effectively transmit the exact color of the avatar model in wireless communication, whereas the color value in the traditional point cloud framework totally up to the channel conditions and will exhibit distortions through wireless communication. 

Fig. \ref{ex44} plots the transmission latency of all frameworks as defined in Eq. (\ref{time_delay}). A lower latency could contribute to a better QoE on the client side, which is ranked as E-GSAR $\textless$ EC-GSAR $\textless$ GSAR $\textless$ Point Cloud. Compared to the traditional point cloud communication framework, the GSAR, E-GSAR, and EC-GSAR save a substantial amount of transmission time due to significantly fewer packets transmitted. Although these frameworks introduce an additional semantic information extraction step with the DL-based semantic information extractor, it only takes about one second per 100 frames, constituting only a tiny portion of the total transmission time.
Concerning pose recovery and rendering, which are inherently linked to the data packets, the point cloud requires rendering all the upsampled point cloud data based on 2,048 points. Conversely, the GSAR, E-GSAR, and EC-GSAR merely require 25 skeletal points to update the pose of an already rendered avatar, thereby significantly reducing time consumption on the client side. Moreover, although both E-GSAR and EC-GSAR necessitate calculating the skeletal position according to Eq. (\ref{eq_l11}) and Eq. (\ref{eq_l12}) before avatar pose recovery, while the GSAR can directly update the avatar pose. The limited calculation time of 25 cycles renders the time consumption of this pose recovery and rendering process relatively uniform among GSAR, E-GSAR, and EC-GSAR. This substantial reduction in data transmission volume concurrently minimizes bandwidth usage 
% and significantly curtails time 
spent on wireless communication compared with the traditional point cloud framework. 

\section{Conclusion}
This paper has presented a novel goal-oriented semantic communication framework designed to enhance the effectiveness and efficiency of avatar-based communication in wireless AR applications. 
By introducing new semantic information in AR and representing relationships between different  types of semantic information using  a graph, our proposed goal-oriented semantic communication framework extracted and transmitted only essential semantic information in wireless AR communication, substantially reducing communication bandwidth requirements. This selective transmission of important semantic information provided a more effective approach to semantic information extraction compared to traditional communication frameworks, ensuring minimal errors and lower bandwidth usage. 
Furthermore, we have extracted effectiveness level features from the complete avatar skeleton graph using shared base knowledge based on end-to-end wireless communication, distinguishing it from and enhancing general semantic communication frameworks. This pioneering work opened research for further advancements in wireless AR communication frameworks.
\textcolor{black}{Our future work will focus on optimizing channel methods, including model recognition, interaction, and channel information estimation. We will also compare the GSAR framework under conditions of both perfect and imperfect channel state information (CSI) feedback by integrating CSI estimation methods, such as pilot-based channel techniques and machine learning algorithms, to further enhance the effectiveness and efficiency of avatar-centric wireless AR applications.}
\ifCLASSOPTIONcaptionsoff
  \newpage
\fi



%\begin{thebibliography}{1}

\small
\bibliographystyle{IEEEtran}
\bibliography{IEEEabrv}

\begin{thebibliography}{10}
\providecommand{\url}[1]{#1}
\csname url@rmstyle\endcsname
\providecommand{\newblock}{\relax}
\providecommand{\bibinfo}[2]{#2}
\providecommand\BIBentrySTDinterwordspacing{\spaceskip=0pt\relax}
\providecommand\BIBentryALTinterwordstretchfactor{4}
\providecommand\BIBentryALTinterwordspacing{\spaceskip=\fontdimen2\font plus
\BIBentryALTinterwordstretchfactor\fontdimen3\font minus \fontdimen4\font\relax}
\providecommand\BIBforeignlanguage[2]{{%
\expandafter\ifx\csname l@#1\endcsname\relax
\typeout{** WARNING: IEEEtran.bst: No hyphenation pattern has been}%
\typeout{** loaded for the language `#1'. Using the pattern for}%
\typeout{** the default language instead.}%
\else
\language=\csname l@#1\endcsname
\fi
#2}}
\renewcommand\BIBentryALTinterwordstretchfactor{4}

\bibitem{ning2021survey}
H.~Ning, H.~Wang, Y.~Lin, W.~Wang, S.~Dhelim, F.~Farha, J.~Ding, and M.~Daneshmand, ``A survey on metaverse: the state-of-the-art, technologies, applications, and challenges,'' \emph{arXiv preprint arXiv:2111.09673}, Nov. 2021.

\bibitem{park2022metaverse}
Y.~Wang, Z.~Su, N.~Zhang, R.~Xing, D.~Liu, T.~H. Luan, and X.~Shen, ``A survey on metaverse: Fundamentals, security, and privacy,'' \emph{IEEE Commun. Surveys Tuts.}, vol.~25, no.~1, pp. 319--352, Sept, 2022.

\bibitem{hu2020cellular}
F.~Hu, Y.~Deng, W.~Saad, M.~Bennis, and A.~Aghvami, ``Cellular-connected wireless virtual reality: Requirements, challenges, and solutions,'' \emph{IEEE Commun. Mag.}, vol.~58, no.~5, pp. 105--111, 2020.

\bibitem{hu2021vision}
F.~Hu, Y.~Deng, H.~Zhou, T.~Jung, C.-B. Chae, and A.~Aghvami, ``A vision of an xr-aided teleoperation system toward \text{5G}/\text{B5G},'' \emph{IEEE Commun. Mag.}, vol.~59, no.~1, pp. 34--40, 2021.

\bibitem{van2020human}
S.~Van~Damme, M.~T. Vega, and F.~De~Turck, ``Human-centric quality management of immersive multimedia applications,'' in \emph{Proc. IEEE Conf. Netw. Softwarization (NetSoft)}, June 2020, pp. 57--64.

\bibitem{dong2022metaverse}
H.~Dong and J.~S. Lee, ``The metaverse from a multimedia communications perspective,'' \emph{IEEE MultiMedia}, vol.~29, no.~4, pp. 123--127, Oct. 2022.

\bibitem{kumar20235g}
M.~N. Kumar, ``\text{5G} technology is revolutionizing the wireless industry with unparalleled efficiency,'' \emph{Sci. Wave. Bull.}, vol.~1, no.~3, pp. 21--28, May 2023.

\bibitem{liao2022kitti}
Y.~Liao, J.~Xie, and A.~Geiger, ``Kitti-360: A novel dataset and benchmarks for urban scene understanding in \text{2D} and \text{3D},'' \emph{IEEE Trans. Pattern Anal. Mach. Intell.}, vol.~45, no.~3, pp. 3292--3310, 2022.

\bibitem{kountouris2021semantics}
M.~Kountouris and N.~Pappas, ``Semantics-empowered communication for networked intelligent systems,'' \emph{IEEE Commun. Mag.}, vol.~59, no.~6, pp. 96--102, June 2021.

\bibitem{yan2022resource}
L.~Yan, Z.~Qin, R.~Zhang, Y.~Li, and G.~Y. Li, ``Resource allocation for text semantic communications,'' \emph{IEEE Wireless Commun. Lett.}, vol.~11, no.~7, pp. 1394--1398, Apr. 2022.

\bibitem{weng2021semantic}
Z.~Weng, Z.~Qin, and G.~Y. Li, ``Semantic communications for speech signals,'' in \emph{Proc. IEEE Int. Conf. Commun. (ICC)}.\hskip 1em plus 0.5em minus 0.4em\relax IEEE, June 2021, pp. 1--6.

\bibitem{jiang2022wireless}
P.~Jiang, C.-K. Wen, S.~Jin, and G.~Y. Li, ``Wireless semantic communications for video conferencing,'' \emph{” IEEE J. Sel. Areas Commun.}, vol.~41, no.~1, pp. 230--244, Nov. 2022.

\bibitem{maatouk2022age}
A.~Maatouk, M.~Assaad, and A.~Ephremides, ``The age of incorrect information: An enabler of semantics-empowered communication,'' \emph{IEEE Trans. Commun.}, Oct. 2022.

\bibitem{zhou2022task}
H.~Zhou, X.~Deng, Yansha~Liu, , N.~Pappas, and A.~Nallanathan, ``Goal-oriented and semantics-aware 6\text{G} networks,'' \emph{arXiv preprint arXiv:2210.09372}, Oct. 2022.

\bibitem{wu2023task}
W.~Wu, Y.~Yang, Y.~Deng, and A.~H. Aghvami, ``Goal-oriented semantics-aware communications for robotic waypoint transmission: the value and age of information approach,'' \emph{arXiv preprint arXiv:2312.13182}, 2023.

\bibitem{du2022optimal}
H.~Du, D.~Niyato, C.~Miao, J.~Kang, and D.~I. Kim, ``Optimal targeted advertising strategy for secure wireless edge metaverse,'' in \emph{Proc. IEEE Global Commun. Conf. (GLOBECOM)}.\hskip 1em plus 0.5em minus 0.4em\relax IEEE, 2022, pp. 4346--4351.

\bibitem{fernandez2022life}
C.~B. Fernandez and P.~Hui, ``Life, the metaverse and everything: An overview of privacy, ethics, and governance in metaverse,'' in \emph{Proc. IEEE Int. Conf. Distrib. Comput. Syst. Workshops (ICDCSW)}.\hskip 1em plus 0.5em minus 0.4em\relax IEEE, July 2022, pp. 272--277.

\bibitem{pauw2022avatar}
L.~S. Pauw, D.~A. Sauter, G.~A. van Kleef, G.~M. Lucas, J.~Gratch, and A.~H. Fischer, ``The avatar will see you now: Support from a virtual human provides socio-emotional benefits,'' \emph{Comput. Human Behav.}, vol. 136, p. 107368, May 2022.

\bibitem{lemmens2023caught}
J.~S. Lemmens and I.~A. Weergang, ``Caught them all: Gaming disorder, motivations for playing and spending among core pok{'e}mon go players,'' \emph{Entertain. Comput.}, p. 100548, March 2023.

\bibitem{da2019point}
L.~A. da~Silva~Cruz, E.~Dumi{'c}, E.~Alexiou, J.~Prazeres, R.~Duarte, M.~Pereira, A.~Pinheiro, and T.~Ebrahimi, ``Point cloud quality evaluation: Towards a definition for test conditions,'' in \emph{Proc. IEEE Int. Conf. Quality of Multimedia Experience (QoMEX)}.\hskip 1em plus 0.5em minus 0.4em\relax IEEE, June 2019, pp. 1--6.

\bibitem{yang2022no}
Q.~Yang, Y.~Liu, S.~Chen, Y.~Xu, and J.~Sun, ``No-reference point cloud quality assessment via domain adaptation,'' in \emph{Proc. IEEE/CVF Conf. Comput. Vision Pattern Recognit.}, 2022, pp. 21\,179--21\,188.

\bibitem{lazzarotto2022influence}
D.~Lazzarotto, M.~Testolina, and T.~Ebrahimi, ``Influence of spatial rendering on the performance of point cloud objective quality metrics,'' in \emph{Proc. 10th European Workshop on Visual Information Processing (EUVIP)}.\hskip 1em plus 0.5em minus 0.4em\relax IEEE, 2022, pp. 1--6.

\bibitem{pavllo20193d}
D.~Pavllo, C.~Feichtenhofer, D.~Grangier, and M.~Auli, ``\text{3D} human pose estimation in video with temporal convolutions and semi-supervised training,'' in \emph{Proc. IEEE/CVF Conference on Computer Vision and Pattern Recognition (CVPR)}, June 2019, pp. 7753--7762.

\bibitem{yu2021avatars}
K.~Yu, G.~Gorbachev, U.~Eck, F.~Pankratz, N.~Navab, and D.~Roth, ``Avatars for teleconsultation: Effects of avatar embodiment techniques on user perception in \text{3D} asymmetric telepresence,'' \emph{IEEE Trans. Vis. Comput. Graph.}, vol.~27, no.~11, pp. 4129--4139, 2021.

\bibitem{xu2021epes}
Y.~Xu, Q.~Yang, L.~Yang, and J.-N. Hwang, ``Epes: Point cloud quality modeling using elastic potential energy similarity,'' \emph{IEEE Trans. Broadcasting}, vol.~68, no.~1, pp. 33--42, 2021.

\bibitem{liu2022deep}
J.~Liu, N.~Akhtar, and A.~Mian, ``Deep reconstruction of \text{3D} human poses from video,'' \emph{IEEE Trans. Artif. Intell.}, pp. 1--1, March 2022.

\bibitem{aseeri2021influence}
S.~Aseeri and V.~Interrante, ``The influence of avatar representation on interpersonal communication in virtual social environments,'' \emph{IEEE transactions on visualization and computer graphics}, vol.~27, no.~5, pp. 2608--2617, 2021.

\bibitem{haruna2023augmented}
M.~Haruna, M.~Ogino, S.~Tagashira, and S.~Morita, ``Augmented avatar toward both remote communication and manipulation tasks,'' in \emph{Proc. IEEE/RSJ Int. Conf. Intelligent Robots and Systems (IROS)}.\hskip 1em plus 0.5em minus 0.4em\relax IEEE, 2023, pp. 7075--7081.

\bibitem{wu2019towards}
Y.~Wu, Y.~Wang, S.~Jung, S.~Hoermann, and R.~W. Lindeman, ``Towards an articulated avatar in vr: Improving body and hand tracking using only depth cameras,'' \emph{Entertain. Comput.}, vol.~31, p. 100303, 2019.

\bibitem{you2020keypointnet}
Y.~You, Y.~Lou, C.~Li, Z.~Cheng, L.~Li, L.~Ma, C.~Lu, and W.~Wang, ``Keypointnet: A large-scale \text{3D} keypoint dataset aggregated from numerous human annotations,'' in \emph{Proc. IEEE/CVF Conf. Computer Vision and Pattern Recognition (CVPR)}, June 2020, pp. 13\,647--13\,656.

\bibitem{zhang2021towards}
Z.-L. Zhang, U.~K. Dayalan, E.~Ramadan, and T.~J. Salo, ``Towards a software-defined, fine-grained \text{QoS} framework for \text{5G} and beyond networks,'' in \emph{Proc. ACM SIGCOMM Workshop Netw.-Appl. Integr. (NAI)}, Aug. 2021, pp. 7--13.

\bibitem{mehrabi2021multi}
A.~Mehrabi, M.~Siekkinen, T.~K{\"a}m{\"a}r{\"a}inen, and A.~Yl{\"a}-J{\"a}{\"a}ski, ``Multi-tier \text{CloudVR}: Leveraging edge computing in remote rendered virtual reality,'' \emph{ACM Trans. Multimed. Comput. Commun. Appl.}, vol.~17, no.~2, pp. 1--24, July 2021.

\bibitem{huang2022iscom}
Y.~Huang, B.~Bai, Y.~Zhu, X.~Qiao, X.~Su, and P.~Zhang, ``Iscom: Interest-aware semantic communication scheme for point cloud video streaming,'' \emph{arXiv preprint arXiv:2210.06808}, Oct. 2022.

\bibitem{nardo2022point}
F.~Nardo, D.~Peressoni, P.~Testolina, M.~Giordani, and A.~Zanella, ``Point cloud compression for efficient data broadcasting: A performance comparison,'' in \emph{Proc. IEEE Wireless Communications and Networking Conference (WCNC)}.\hskip 1em plus 0.5em minus 0.4em\relax IEEE, March 2022, pp. 2732--2737.

\bibitem{qi2017pointnet++}
C.~R. Qi, L.~Yi, H.~Su, and L.~J. Guibas, ``Pointnet++: Deep hierarchical feature learning on point sets in a metric space,'' \emph{Adv. Neural Inf. Process. Syst.}, vol.~30, Dec. 2017.

\bibitem{akhtar2022pu}
A.~Akhtar, Z.~Li, G.~Van~der Auwera, L.~Li, and J.~Chen, ``{Pu-Dense}: Sparse tensor-based point cloud geometry upsampling,'' \emph{IEEE Trans. Image Process.}, vol.~31, pp. 4133--4148, July 2022.

\bibitem{chen2020pointmixup}
Y.~Chen, V.~T. Hu, E.~Gavves, T.~Mensink, P.~Mettes, P.~Yang, and C.~G. Snoek, ``Pointmixup: Augmentation for point clouds,'' in \emph{Proc. Eur. Conf. Comput. Vis. (ECCV)}.\hskip 1em plus 0.5em minus 0.4em\relax Springer, June 2020, pp. 330--345.

\bibitem{egilmez2019development}
Z.~B.~K. Egilmez, L.~Xiang, R.~G. Maunder, and L.~Hanzo, ``Development, operation, and performance of \text{5G} polar codes,'' \emph{IEEE Commun. Surv. Tutor.}, vol.~22, no.~1, pp. 96--122, 2019.

\bibitem{quintero2022excite}
L.~Quintero, P.~Papapetrou, J.~E. Mu{~n}oz, J.~De~Mooij, and M.~Gaebler, ``Excite-o-meter: an open-source unity plugin to analyze heart activity and movement trajectories in custom vr environments,'' in \emph{2022 IEEE Conf. Virtual Reality \text{3D} User Interfaces Abstracts Workshops (VRW)}.\hskip 1em plus 0.5em minus 0.4em\relax IEEE, 2022, pp. 46--47.

\bibitem{thoota2022massive}
S.~S. Thoota and C.~R. Murthy, ``Massive \text{MIMO-OFDM} systems with low resolution adcs: Cram{'e}r-rao bound, sparse channel estimation, and soft symbol decoding,'' \emph{IEEE Trans. Signal Process.}, vol.~70, pp. 4835--4850, 2022.

\bibitem{qiu2021dense}
S.~Qiu, S.~Anwar, and N.~Barnes, ``Dense-resolution network for point cloud classification and segmentation,'' in \emph{Proc. IEEE/CVF Winter Conf. on Applications of Computer Vision (WACV)}.\hskip 1em plus 0.5em minus 0.4em\relax IEEE, 2021, pp. 3813--3822.

\bibitem{joshi2018google}
M.~A. Joshi and P.~Patel, ``Google page rank algorithm and it's updates,'' in \emph{Proc. Int. Conf. Emerg. Trends Sci. Eng. Manage.(ICETSEM)}, 2018.

\bibitem{srivastava2017discussion}
A.~K. Srivastava, R.~Garg, and P.~Mishra, ``Discussion on damping factor value in pagerank computation,'' \emph{Int. J. Intell. Syst. Appl.}, vol.~9, no.~9, p.~19, Sept. 2017.

\bibitem{mekuria2016evaluation}
R.~Mekuria, Z.~Li, C.~Tulvan, and P.~Chou, ``Evaluation criteria for pcc (point cloud compression),'' ISO/IEC JTC1/SC29/WG11, Technical Report n16332, June 2016.

\bibitem{meynet2020pcqm}
G.~Meynet, Y.~Nehm{\'e}, J.~Digne, and G.~Lavou{\'e}, ``Pcqm: A full-reference quality metric for colored \text{3D} point clouds,'' in \emph{2020 Twelfth International Conference on Quality of Multimedia Experience (QoMEX)}.\hskip 1em plus 0.5em minus 0.4em\relax IEEE, 2020, pp. 1--6.

\bibitem{fiedler2010generic}
M.~Fiedler, T.~Hossfeld, and P.~Tran-Gia, ``A generic quantitative relationship between quality of experience and quality of service,'' \emph{IEEE Netw.}, vol.~24, no.~2, pp. 36--41, March 2010.

\end{thebibliography}

%\end{thebibliography}
% \begin{biography}[{\includegraphics[width=1in,height=1.25in,clip,keepaspectratio]{mshell}}]{Michael Shell}

% ==== SWITCH OFF the BIO for submission
% ==== SWITCH OFF the BIO for submission
\begin{IEEEbiography}[{\includegraphics[width=1in,height=1.25in,clip,keepaspectratio]{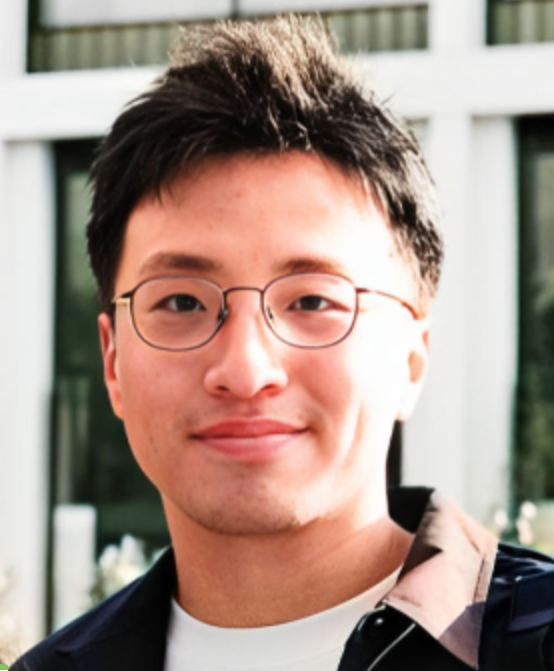}}]{Zhe Wang}
(Graduate Student Member, IEEE) is currently pursuing the Ph.D. degree with the Center for Telecommunications Research (CTR), King’s College London. His current research interests include wireless semantic communication for metaverse applications. 
\end{IEEEbiography}
\begin{IEEEbiography}[{\includegraphics[width=1in,height=1.25in,clip,keepaspectratio]{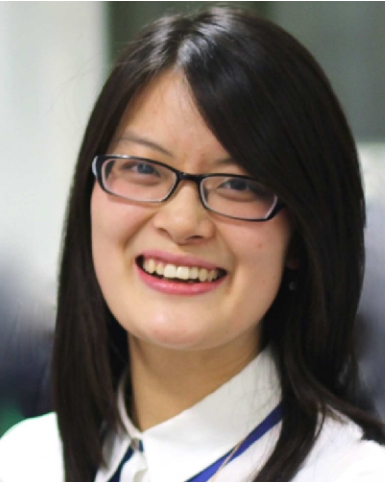}}]{Yansha Deng}
(Senior Member, IEEE) received the Ph.D. degree in electrical engineering from the Queen Mary University of London, U.K., in 2015. From 2015 to 2017, she was a Post-Doctoral Research Fellow with King’s College London, U.K., where she is currently a Reader (an Associate Professor) with the Department of Engineering. Her current research interests include molecular communication and machine learning for 5G/6G wireless networks. She was a recipient of the Best Paper Awards from ICC in 2016 and GLOBECOM in 2017 as the first author and IEEE Communications Society Best Young Researcher Award for the Europe, Middle East, and Africa Region in 2021. She also received the exemplary reviewers of the IEEE TRANSACTIONS ON COMMUNICATIONS in 2016 and 2017 and IEEE TRANSACTIONS ON WIRELESS COMMUNICATIONS in 2018. She has also served as a TPC Member for many IEEE conferences, such as IEEE GLOBECOM and ICC. She is currently an Associate Editor of IEEE TRANSACTIONS ON COMMUNICATIONS, IEEE COMMUNICATIONS SURVEYS AND TUTORIALS, IEEE TRANSACTIONS ON MACHINE LEARNING IN COMMUNICATIONS AND NETWORKING, and IEEE TRANSACTIONS ON MOLECULAR, BIOLOGICAL AND MULTI-SCALE COMMUNICATIONS; a Senior Editor of the IEEE COMMUNICATION LETTERS; and the Vertical Area Editor of IEEE Internet of Things Magazine.
\end{IEEEbiography}
\begin{IEEEbiography}[{\includegraphics[width=1in,height=1.25in,clip,keepaspectratio]{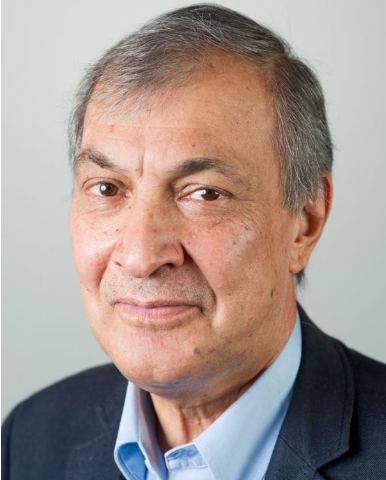}}]{A. Hamid Aghvami}
(Life Fellow, IEEE) joined King’s College London, an Academic Staff, in 1984. In 1989, he was promoted to a Reader, and was promoted to a Professor of telecommunications engineering in 1993. He was a Visiting Professor at NTT Radio Communication Systems Laboratories in 1990, a Senior Research Fellow at BT Laboratories from 1998 to 1999, and was an Executive Advisor of Wireless Facilities Inc., USA, from 1996 to 2002. He was the Director of the Centre from 1994 to 2014. He is currently the Founder of the Centre for Telecommunications Research, King’s College London. He is the Chairperson of Advanced Wireless Technology Group Ltd. He is also the Managing Director of Wireless Multimedia Communications Ltd., his own consultancy company. He is also a Visiting Professor at Imperial College London. He carries out consulting work on digital radio communications systems for British and international companies. He has published over 580 technical journals and conference papers, filed 30 patents, and given invited talks and courses the world over on various aspects of mobile radio communications. He leads an active research team working on numerous mobile and personal communications projects for fourth and fifth generation networks; these projects are supported both by government and industry. He was a member of the Board of Governors of the IEEE Communications Society from 2001 to 2003, was a Distinguished Lecturer of the IEEE Communications Society from 2004 to 2007, and has been a member, the Chairperson, and the Vice-Chairperson of the technical program and organizing committees of a large number of international conferences. He is a fellow of the Royal Academy of Engineering and a fellow of the IET. He was awarded the IEEE Technical Committee on Personal Communications (TCPC) Recognition Award in 2005 for his outstanding technical contributions to the communications field, and for his service to the scientific and engineering communities. In 2009, he was awarded a Fellowship of the Wireless World Research Forum in recognition of his personal contributions to the wireless world, and for his research achievements as the Director of the Centre for Telecommunications Research, King’s College London. He is also the Founder of the International Symposium on Personal Indoor and Mobile Radio Communications (PIMRC), a major yearly conference attracting some 1,000 attendees.
\end{IEEEbiography}

\vfill


\end{document}